\documentclass[aps,prl,a4paper,twocolumn, showkeys]{revtex4}

\usepackage[]{amsmath}
\usepackage{textcomp,wasysym}
\usepackage{amssymb}
\usepackage[caption = false]{subfig}
\usepackage[dvips]{graphicx}
\usepackage{color}
\usepackage{tabularx}
\usepackage{mathtools}
\usepackage{algpseudocode}
\usepackage{enumitem}
\usepackage{multirow}
\usepackage{epstopdf}  
\usepackage{hyperref}
\usepackage{footnote}
\usepackage[hang,flushmargin]{footmisc}
\usepackage{bm}
\usepackage{xr}

\makeatletter
\newcommand*{\rom}[1]{\expandafter\@slowromancap\romannumeral #1@}
\makeatother

\newcommand{\dft}{\mathrm{DFT}}
\newcommand{\deep}{\mathrm{DP}}
\newcommand{\dpname}[0]{DP}
\newcommand{\meam}{\mathrm{MEAM}}
\newcommand{\voigt}{\mathrm{V}}

\newcommand{\aform}{\mathrm{af}}

\newcommand{\al}{{\mathrm{Al}}}
\newcommand{\mg}{{\mathrm{Mg}}}
\newcommand{\me}{\mathcal {E}}

\newcommand{\methodname}[0]{DP-GEN}

\usepackage{booktabs}
\usepackage{siunitx}
\newcommand{\beginsupplement}{%
  \setcounter{table}{0}
  \renewcommand{\thetable}{S\arabic{table}}%
  \setcounter{figure}{0}
  \renewcommand{\thefigure}{S\arabic{figure}}%
}

\begin{document}

\title{
Active Learning of Uniformly Accurate Inter-atomic Potentials for Materials Simulation
}
\author{Linfeng Zhang}
\affiliation{Program in Applied and Computational Mathematics, Princeton University, Princeton, NJ 08544, USA}
\author{De-Ye Lin}
\affiliation{Institute of Applied Physics and Computational Mathematics, Huayuan Road 6, Beijing 100088, P.R.~China}
\affiliation{CAEP Software Center for High Performance Numerical Simulation, Huayuan Road 6, Beijing 100088, P.R.~China}
\author{Han Wang}
\email{wang\_han@iapcm.ac.cn}
\affiliation{Laboratory of Computational Physics, Institute of Applied Physics and Computational Mathematics, Huayuan Road 6, Beijing 100088, P.R.~China}
\author{Roberto Car}
\affiliation{Department of Chemistry,  
Department of Physics, 
Program in Applied and Computational Mathematics, 
Princeton Institute for the Science and Technology of Materials,  
Princeton University, Princeton, NJ 08544, USA}
\author{Weinan E}
\email{weinan@math.princeton.edu}
\affiliation{Department of Mathematics and Program in Applied and Computational Mathematics, Princeton University, Princeton, NJ 08544, USA}
\affiliation{Beijing Institute of Big Data Research, Beijing, 100871, P.R.~China}

\begin{abstract}
An active learning procedure called Deep Potential Generator (DP-GEN) is proposed for the construction of accurate and transferable machine learning-based models of the potential energy surface (PES) for the molecular modeling of materials. 
This procedure consists of three main components: exploration, generation of accurate reference data, and training. 
Application to the sample systems of Al, Mg and Al-Mg alloys demonstrates that DP-GEN can produce uniformly accurate PES models with a minimal number of reference data.
\end{abstract}

\maketitle
\section{Introduction}
The inter-atomic potential energy surface (PES) plays a central role in the molecular modeling of materials.
Obtaining an accurate and efficient representation of the PES is a central issue in molecular simulation.
In this context, one faces the dilemma that {\it ab initio} methods are accurate but highly inefficient, while empirical force fields (FFs) are efficient, but there is a limited guarantee for their accuracy.
Thus, there is a great demand for an efficient and uniformly accurate PES model that can be used to compute a broad range of atomistic properties for most material compounds of practical interest.

Developing empirical FFs has been challenging due to the high dimensionality and many-body character of the PES.
Usually, empirical FFs parameterize the PES by assuming an analytical functional form in terms of relatively simple functions based on physical/chemical intuition, and by fitting the model parameters against a bundle of experimental properties and/or microscopic quantities from \emph{ab initio} calculations.
Some popular examples are the Lennard-Jones potential~\cite{jones1924LJ}, the Stillinger-Weber potential~\cite{stillinger1985SW}, the embedded-atom method (EAM) potential~\cite{daw1984EAM}, the CHARMM~\cite{mackerell1998charmm}/AMBER~\cite{wang2000amber} FFs, the reactive FFs~\cite{van2001rff}, etc. 
\emph{Representability} and \emph{transferability} are two main issues faced by empirical FFs.
By representability, we mean the ability of the assumed functional form to reproduce accurately the target properties.
By transferability, we mean the ability of a PES model to describe properties that do not belong to the set of fitting targets.
Due to the physical/chemical knowledge encoded in the functional form, we expect the empirical FFs to be qualitatively transferable to a moderate range of thermodynamic conditions beyond those adopted for the fitting.
However, as a consequence of assuming relatively simple functional forms, empirical FFs usually face a severe representability problem.
Moreover, a substantial human effort in tuning the model parameters is often required to achieve the best balance in fitting the target properties. 

Recent progress with machine learning (ML) methods is changing the outlook~\cite{behler2007generalized, bartok2010gaussian,rupp2012fast,montavon2013machine,botu2016machine,chmiela2017machine,schutt2017schnet,bartok2017machine,smith2017ani,han2017deep,chen2018atomic,zhang2018deep,zhang2018end}.
ML models, being capable of learning complex and highly nonlinear functional dependence, are excellent in their representability.
It is now possible, using modern ML approaches, to parametrize the PES using data from {\it ab initio} calculations to obtain models that have {\it ab initio} accuracy and are, at the same time, competitive regarding efficiency against empirical FFs. 
In spite of the remarkable success of these ML methods, there is no guarantee for the quality of ML models when they are used to predict the properties of a configuration that is far from the training dataset~\cite{lecun2015deep}.
In addition, since the training data is usually generated with expensive first-principle calculations, one would like to obtain good ML models without having to rely on very large {\it ab initio} datasets.
These questions arise not only for PES modeling, but in many other contexts when ML methods are applied to problems involving physical models.

To address this issue, we get inspiration from active learning~\cite{settles2012active,rubens2015active}, an area of supervised learning whose aim is to learn general purpose models with a minimal number of training data. 
A training data point involves an input and an output.
For example, in an image recognition task whose goal is to judge whether a cat is in an image or not, the input is an array of digits that represents the image, and the output is a boolean proposition.
Usually the output is called a label and the term labeling is used to denote the creation of a label.
In the context of active learning, one typically faces a situation in which unlabeled data are abundant, but labeling is expensive.
Therefore, an interactive algorithm is required to efficiently explore unlabeled data, collect feedbacks on-the-fly, and actively query the teacher for labels on data points with negative feedbacks.
Along this line of thinking, at an abstract level, one can formulate an active learning procedure for PES modeling that involves three steps: exploration, labeling, and training.
\begin{enumerate}
\item \emph{Exploration} requires an efficient \emph{sampler} and an informative \emph{indicator}. 
The sampler uses the current PES model to quickly explore the configuration space.
The indicator monitors on-the-fly the configurations explored by the sampler, selects those with low prediction accuracy, and sends them to the labeling step.
\item \emph{Labeling} means generating reference {\it ab initio} energies and forces for the selected configurations.
Labeling can be done by a code that implements high-level quantum chemistry, quantum Monte Carlo, or density functional theory (DFT) methods.
The labeled configurations are then \emph{added} to the existing dataset and used in the new iteration for training.
\item \emph{Training} requires a good \emph{model}, or PES representation, which can fit the ever-increasing dataset with satisfactory accuracy.
Such a representation should be efficient and should satisfy certain physical constraints like the extensive and symmetry-preserving properties of the PES.
\end{enumerate}
The whole scheme falls into a closed loop: One starts with a relatively poor approximation of the PES and uses it to explore different configurations.
Then a selected set of new configurations is labeled, and a new approximation of the PES is obtained by training.
These three steps are repeated until convergence is achieved, i.e.,
the configuration space has been explored sufficiently, and a minimal set of data points have been accurately labeled.
At the end of this procedure, a uniformly accurate PES model is generated.

In this work, our first goal is to translate the general proposal described above into a practical scheme for modeling the PES.
In this scheme, for the PES representation, we use an advanced version of the Deep Potential (DP) model~\cite{zhang2018end}, which has shown great promise in learning the PES of a broad range of systems, such as insulators, molecular crystals, and a 5-component high entropy alloy, etc.
See, e.g., Fig. 1 of Ref.~\cite{zhang2018end}.
For the sampler, we use molecular dynamics (MD) based on the DP model. Thereafter DP based MD will be referred to as DPMD.
At the same time, we introduce an indicator that we call the model deviation.
This is done as follows. We train an ensemble of \dpname{}~models using the same dataset but different initialization of the DP parameters.
For each new configuration that is explored by DPMD, these models generate an ensemble of predictions.
For each configuration, the model deviation is defined as the maximum standard deviation of the predicted atomic forces. 
A high model deviation indicates low quality in the model prediction and is proposed for labeling.
In this work, we use in the labeling stage DFT within the generalized gradient approximation~\cite{perdew1996generalized,kresse1996efficient,kresse1996efficiency,monkhorst1976special}, which works well in the chosen testing examples.
We will see that sampling is much cheaper than labeling, and only a very small fraction of the explored configurations is selected for labeling. 
We call the methodology introduced here the Deep Potential Generator, abbreviated~\methodname{}.

Our second goal is to demonstrate the uniform accuracy of a PES model obtained in this way.
To this end, we consider the example of Al, Mg, as well as Al-Mg alloys.
Using \methodname{}, we construct a model that can accurately describe these systems at different compositions and thermodynamic conditions.
The resulting PES model is evaluated from the point of view of a material scientist.
We calculate several statical, dynamical, and mechanical properties, such as radial distribution functions (RDF), phonon spectra, elastic constants, etc.
Some of these properties are compared with DFT results.  
We also compare DP calculated properties directly with experimental results when these are available.
To further test the quality of the PES model, we introduce an automatic procedure based on the Materials Project (MP) database~\cite{jain2013commentary}.
In this procedure, one searches the database by entering a material composition, such as Al-Mg in the present case.
The database will then return a large number of locally stable structures,  including many structures of potential practical interest.
Based on these structures, we evaluate several equilibrium properties and compare the DFT predictions with those of the PES model.
In addition, for each one of these structures, we automatically generate unrelaxed vacancy and interstitial defects as well as the set of surfaces corresponding to a range of Miller indices~\cite{ong2013python}.
We then compare the relaxed formation energies of the defects and the unrelaxed formation energies of the surfaces predicted by DFT and by the PES model.
We stress that these structures, i.e., crystals, defects, and surfaces, were not explicitly included in the training data.
We find that our PES model can achieve uniform accuracy in the prediction of all of these structural properties.

We notice that there is a difference between active learning in conventional ML problems and the active learning we pursue here.
This difference lies in exploration or sampling.
Conventional active learning problems in ML typically deal with an existing unlabeled dataset.
Here our dataset is generated on the fly via sampling. 
This means that we need to have an efficient sampling method.

We should mention that related work can be found in the literature
~\cite{podryabinkin2017active,smith2018less,herr2018metadynamics,bonati2018silicon,musil2019fast}. 
In particular, Smith et al~\cite{smith2018less} utilized an active learning scheme to model the PES of organic molecules based on an existing large database~\cite{smith2017ani-data}. 
Moreover, Bartok et al~\cite{bartok2018machine} constructed a kernel based general purpose PES model for pure silicon, wherein they exhaustively enumerated possible structures for labeling.
Finally, the principle of active learning was also used in the reinforced dynamics scheme~\cite{zhang2018reinforced} for enhanced sampling and free energy calculation.

\section{Methodology}

\begin{figure}[ht!]
  \centering
  \includegraphics[width=0.48\textwidth]{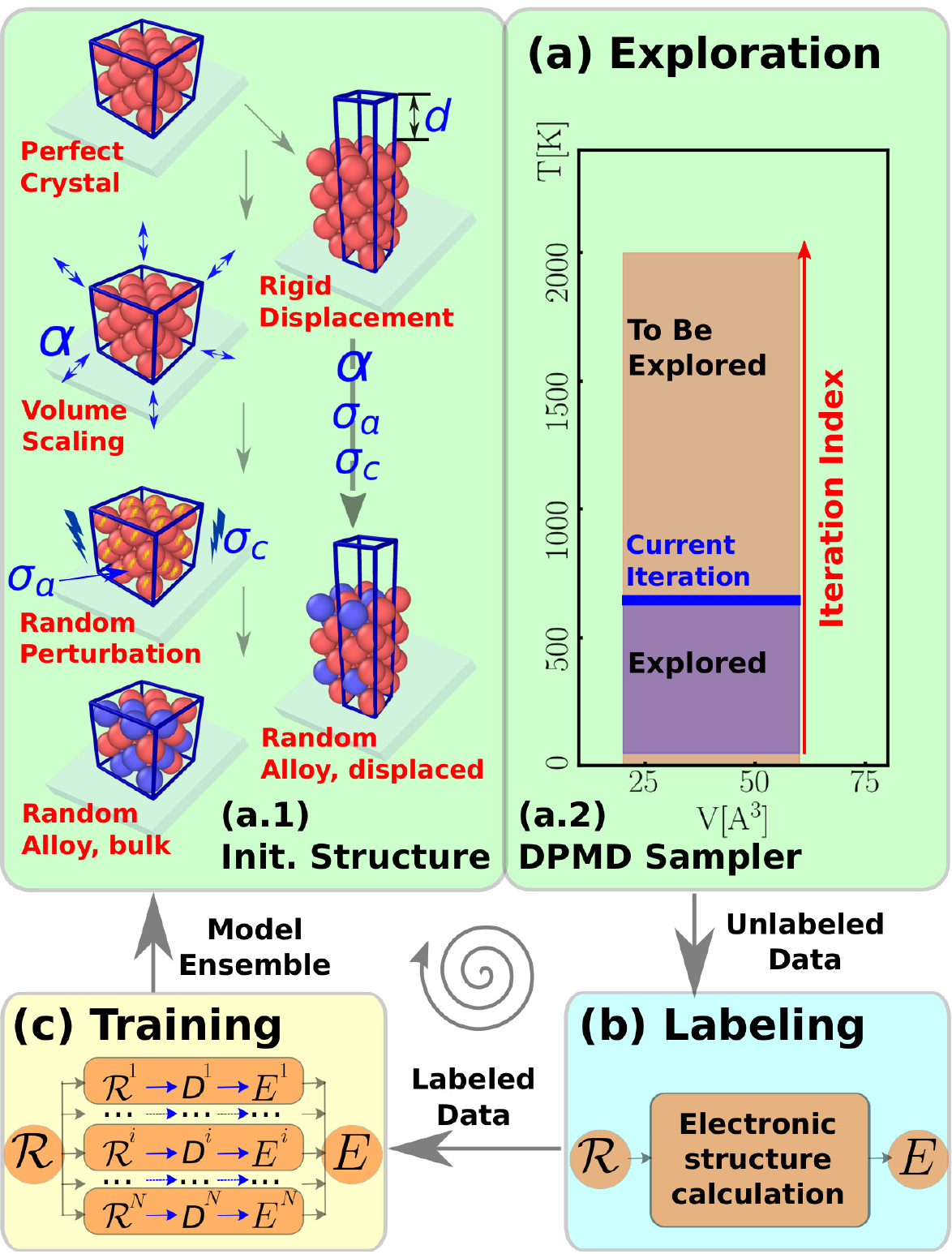}
  \caption{
Schematic plot of one iteration of the \methodname{} scheme, taking the Al-Mg system as an example.
(a) Exploration with DPMD.
(a.1) Preparation of initial structures.
\rom{1}. For bulk structures: start from stable crystalline structures of pure Al and Mg.
In this work, we use face-centered-cubic (FCC), hexagonal-closed-packed (HCP), simple cubic (SC), and diamond structures.
\rom{2}. Compress and dilate the stable structures uniformly to allow for a larger range of number densities. 
We use $\alpha$ to denote the scale factor of the compression and dilation operations. Here $\alpha$ ranges in the interval 0.96-1.04.
\rom{3}. Randomly perturb the atomic positions and cell vectors of all the initial crystalline structures.
The magnitude of perturbations on the atomic coordinates is $\sigma_a=0.01\AA$.
The magnitude of perturbation on each cell vector is $\sigma_c=0.03$ times the length of the cell vector.
\rom{4}. Generate random alloy structures: starting from all the structures prepared for pure systems, randomly place Al or Mg at different sites.
\rom{5}. Generate structures with rigid displacement: starting from stable FCC and HCP structures, rigidly displace two crystalline halves along specific crystallographic directions.
We only use (100), (110), (111), and (0001), (10$\bar{1}$0), (11$\bar{2}$0), respectively, for FCC and HCP, as the displacement directions.
The magnitudes $d$ of the displacements range in the interval 0.2-10.0\AA.
Based on all the displaced structures, perform dilation $\alpha$ and perturbation $\sigma_a$ and $\sigma_c$, and generate random alloy structures.
(a.2) Canonical simulation at a given temperature. The temperature increases with the iteration index within the range 50-2000~K.
(b) Labeling with electronic structure calculations.
(c) Training with the DP model.
}
  \label{fig:method}
\end{figure}

In this section, we introduce the three essential components of the \methodname~scheme: the model, the sampler, and the indicator.
Fig.~\ref{fig:method} shows a schematics of \methodname.
To initialize  the procedure, 
we label a small set of initial structures introduced in Fig.~\ref{fig:method}(a) and train an ensemble of preliminary DP models.
More details on the simulation protocol and the iterative process are reported in the supplementary materials (SM).

{\it Model}. 
The \dpname~scheme assumes that the potential energy $E$ can be written as a sum of atomic energies, i.e., $E=\sum_iE^i$.
Each atomic energy $E^i$ is a function of  $\mathcal R^i$, the local environment of atom $i$ in terms of the relative coordinates of its neighbors within a cut-off radius $r_c$.
The dependence of $E^i$ on $\mathcal R^i$ embodies the nonlinear and many-body character of the inter-atomic interactions.
Therefore, we use a deep neural network function (DNN) to parameterize it, i.e., $E^i=E^{\bm{w}_{\alpha_i}}(\mathcal{R}^i)$.
Here $\alpha_i$ indicates the chemical species of the $i$-th atom; $\bm{w}_{\alpha_i}$ denotes the parameters of the DNN we call network parameters, that are determined by the training procedure.
A vital component of the \dpname~model is a general procedure that encodes $\mathcal R^i$ into the so-called feature matrix $\mathcal D^i$.
This procedure guarantees the conservation of the translational, rotational, and permutational symmetries of the system, without losing coordinate information in the local environment. 
Derivatives of the energy with respect to the atomic positions give the forces.
During the training process, the network parameters evolve in order to minimize the loss function, a measure of the error in the energies and the forces predicted by DP relative to the labels, i.e., the corresponding DFT predictions~\cite{kingma2015adam}.
Upon convergence, the model can match the labels within a small error tolerance.
The details of the architecture of the \dpname~model and the training process are given in Ref.~\cite{zhang2018end}.

{\it Sampler}. 
The goal of the sampler is to explore the configuration space in a range of thermodynamic variables, say temperature and pressure.  
Ideally one should develop an automatic/adaptive procedure for this purpose. 
However, since exploration is relatively cheap compared to labeling, we adopt a more heuristic approach in which
the exploration is done through: 
(1) carefully selecting the initial configurations, and (2) exploring the volume-temperature space.
We use a variety of crystal structures as our initial configuration, as in the procedure illustrated in Fig.~\ref{fig:method}(a).
To explore the volume-temperature phase space, we adopt a temperature increasing scheme, in which the temperature of the DPMD simulations is increased systematically with the iteration index in the range 50-2000~K.
We notice that many structures constructed in this way are far from equilibrium structures so that the subsequent DPMD simulations in the 50-2000K temperature range produce a large sample of configurations that may differ substantially from the initial structure.
More details on the initial structures and the thermodynamic conditions in each iteration are summarized in Tables~\ref{tab:data:02-al}--\ref{tab:data:00}.

{\it Indicator}. 
It is well-known that neural network models are highly nonlinear functions of the network parameters $\bm{w}_{\alpha_i}$.
The loss function, as a function of $\bm{w}_{\alpha_i}$, is highly non-convex, i.e., several local minima exist in the landscape of the loss function.
In the current work, we initialize the $\bm{w}_{\alpha_i}$ randomly according to the standard normal distribution.
As a result, different initializations often lead to different minimizers of the loss function. 
These minimizers fit well the training data, so in the configurational region belonging to the neighborhood of the training data, they generate equally accurate energies and forces and show small deviations in their predictions.
However, for snapshots ``far'' from the training data, these minimizers usually predict inaccurate values that show significantly larger deviation.
This property of neural network models motivates us to define the indicator as the deviation of the predictions generated by an ensemble of \dpname~models trained with the same dataset but with different parameter initializations.
In practice, we define the model deviation, denoted as $\me$, as the maximum standard deviation of the predictions for the atomic forces, i.e.:
\begin{align}\label{eqn:model-std}
  \me =\max_i
  \sqrt{\big\langle
  \Vert\bm{f}_i - \bar{\bm{f}_i}
  \Vert^2
  \big\rangle},
  \quad
  \bar{\bm{f}_i} =
  \big\langle
  \bm{f}_i
  \big\rangle,
\end{align}
where $i$ runs through the atomic indices in a configuration, and the ensemble average $\langle\cdots\rangle$ is taken over the ensemble of models.
We find that using the predicted forces to evaluate the model deviation is generally better than using the predicted energies.
The force is an atomic property and is sensitive to a failure in local predictions, while the energy is a global quantity and does not seem to provide sufficient resolution in this regard.
Moreover, we find that a failure in local predictions can be better signaled by using the maximum over $i$ in Eqn.~\ref{eqn:model-std}, instead of the average over $i$ ($\frac{1}{N}\sum_i$).

\section{Results}
As examples, we report the results of the \methodname~scheme for Al, Mg and their alloys.
At the end of the \methodname~scheme, we collect a set of labeled data and obtain a \dpname~model for the Al-Mg system.
As shown in Table~\ref{tab:data:00}, about 650 million configurations were explored by DPMD, but only 0.0044\% of them were selected for
 labeling.
To get an idea of the usefulness of the resulting \dpname~model for materials science applications, we compare the accuracy of the DP model in predicting important material properties with a state-of-the-art empirical FF like the modified embedded atom method (MEAM)~\cite{baskes1992modified}.
MEAM adopts a more general definition of embedding than EAM in order to improve the description of directional bonding and of alloy systems. 
In this work, we compare our method with a very recent version of the Al-Mg MEAM potential that is available in the literature~\cite{jelinek2012modified}.
We used DeePMD-kit~\cite{wang2018kit} in the training step, LAMMPS~\cite{plimpton1995lammps} in the exploration step, and VASP~\cite{kresse1996efficiency,kresse1996efficient} in the labeling step.

\subsection{Pure Al and Mg}
\begin{figure}
  \centering
  \includegraphics[width=0.45\textwidth]{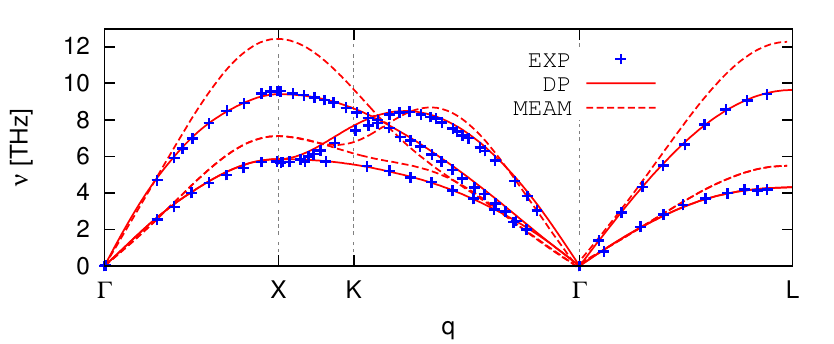}
  \caption{The phonon dispersion relations of Al at $T=80$K and $P=1$bar. Here $q$ denotes the wave number and $\nu$ the frequency. The experimental data is taken from~\cite{stedman1966dispersion}.}
  \label{fig:al:02}
\end{figure}

\begin{figure}
  \centering
  \includegraphics[width=0.45\textwidth]{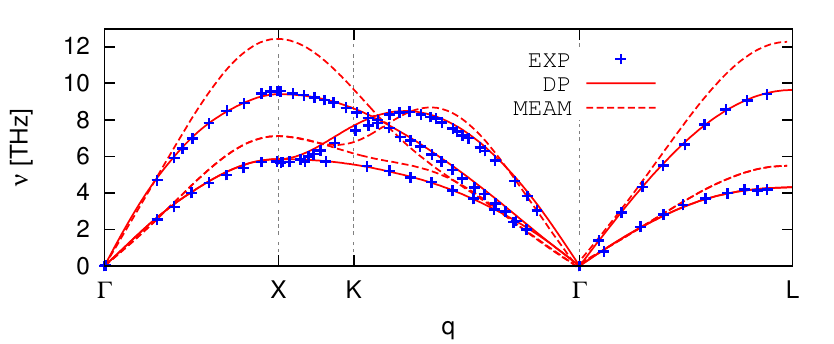}
  \caption{The RDFs $g(r)$ of liquid Al at $P=1$~bar and temperatures $T = 943$~K.
    The DP and MEAM predictions are compared with the experimental data taken from~\cite{waseda1980structure}.
    The inserted plot is the zoom-in of RDFs in range $3.5$\AA$\leq r \leq 7$\AA.
  }
  \label{fig:al:03}
\end{figure}

The equilibrium properties of pure Al are presented in Table~\ref{tab:al:00}, including the atomization energy and equilibrium volume per atom, defect formation energies,
elastic constants and moduli, stacking fault energies, melting point, enthalpy of fusion, and diffusion coefficient.
The defect formation energy is defined as
$
E_{\textrm{df}} = E_{\textrm{d}}(N_{\textrm{d}}) - {N_{\textrm{d}}} E_0, 
$
$d = v(i)$ indicating vacancy (interstitial) defects.
$E_{\textrm{d}}$ denotes the relaxed energy of a defective structure with $N_{\textrm{d}}$ atoms 
and $E_0$ denotes the energy per atom of the corresponding ideal crystal at $T=0$~K.
To compute the defect formation energies, we use a supercell in which we replicate $7\times 7\times 7$ times the primitive FCC cell.
We estimate the melting temperature ($T_m$) by simulating with DPMD coexisting crystal and liquid phases in a supercell containing 8000 atoms within the isothermal-isobaric ensemble at standard pressure.
To estimate the liquid diffusion coefficient ($D$),  we perform DPMD simulations on large supercells (6912 atoms) for which finite size effects are negligible.
For all the properties in Table~\ref{tab:al:00}, the DP predictions are in satisfactory agreement with DFT and/or experiment.
Notice that MEAM reproduces quite accurately the solid state properties in Table~\ref{tab:al:00}, particularly when compared to experiment, which is not surprising since the basic experimental solid state properties have been used to tune the parameters of this FF. 
However, the vibrational properties at short wavelength, particularly the zone boundary phonons, are not reproduced well by MEAM in contrast to DP, as shown in Fig.~\ref{fig:al:02}.
MEAM fails even more dramatically in predicting the properties of the liquid: the MEAM liquid is largely overstructured (see Fig.~\ref{fig:al:03}).
Its diffusion coefficient is one order of magnitude smaller than in experiment or DP, and its enthalpy of fusion is also significantly smaller than in experiment or DP (see Table~\ref{tab:al:00}).

\begin{table}
  \centering
  \caption{Equilibrium properties of Al: atomization energy $E_{\textrm{am}}$, equilibrium volume per atom $V_0$,
  vacancy formation energy $E_{\textrm{vf}}$, interstitial formation energies $E_{\textrm{if}}$ for octahedral interstitial (oh) and tetrahedral interstitial (th).
independent elastic constants $C_{11}$, $C_{12}$, and $C_{44}$, Bulk modulus $B_\voigt$ (Voigt), shear modulus $G_\voigt$ (Voigt),
stacking fault energy $\gamma_{\textrm{sf}}$,
twin stacking fault energy $\gamma_{\textrm{tsf}}$, melting point $T_m$, enthalpy of fusion $\Delta{H}_f$, and diffusion coefficient $D$ at $T=1000$K.}
  \label{tab:al:00}
  \begin{tabular*}{0.48\textwidth}{@{\extracolsep{\fill}} lrrrr}
    \hline\hline   
    Al              & EXP   &DFT\footnote{The DFT results, unless specified with a reference, are computed by the authors.
    We notice that a K-mesh spacing equal to $0.06$~\AA$^{-1}$ was used to obtain more converged DFT results in this table.
    However, in the labeling stage, we used a K-mesh spacing equal to $0.08$~\AA$^{-1}$, which gives converged values for most of the properties except for elastic constants and moduli. 
    Using K-mesh spacing equal to $0.08$~\AA$^{-1}$ gives $C_{11}$ = 129.3~GPa, $C_{12}$ = 52.8~GPa, $C_{44}$ = 37.4~GPa, $B_\voigt$ = 78.3~GPa, and $G_\voigt$ = 37.7~GPa.
    ${}^b$Ref.~\cite{cox1989codata}. 
    ${}^c$Experiment values obtained at $T=298$K; DFT, DP, and MEAM results obtained at $T=0$K.
    ${}^d$Ref.~\cite{cooper1962precise}.
    ${}^e$Refs.~\cite{triftshauser1975positron,fluss1978measurements}.
    ${}^f$Ref.~\cite{hood2012diffusion}.
    ${}^g$Ref.~\cite{kamm1964low}.
    ${}^h$Refs.~\cite{kannan1966dislocation,dobson1967climb,tartour1968climb,mills1989study}.
    ${}^i$Ref.~\cite{zhao2016impurity}.
    ${}^j$Ref.~\cite{ross2004melting}.
    ${}^k$Ref.~\cite{bouchet2009melting}.
    ${}^l$Ref.~\cite{mcdonald1967enthalpy}.
    ${}^m$Ref.~\cite{meyer2015measurement}, $D=7.2\times 10^{-9}$m$^2$/s at 980K and $7.9\times 10^{-9}$m$^2$/s at 1020K. 
    }
    & DP  &MEAM  \\\hline
    $E_{\textrm{am}}$ [eV/atom]      & $-3.49{}^b$
    &$-3.655$ & $-3.654$ & $-3.353$  \\
    $V_0$ [\AA$^3$/atom]${}^c$
    & $16.50$${}^d$
    &$16.48$ &   $16.51$ & $16.61$  \\
    $E_{\textrm{vf}}$ [eV]      & 0.66${}^e$
    &$0.67{}^f$
    &  $0.79$  & 0.67 \\
    $E_{\textrm{if}}$ (oh) [eV] & -     &$2.91{}^f$
    &  $2.45$  &  2.77 \\
    $E_{\textrm{if}}$ (th) [eV] & -     &$3.23{}^f$
    &  $3.12$  &  3.32 \\
    $C_{11}$ [GPa]  & 114.3${}^g$
    & $111.2$ &$120.9$&  111.4 \\
    $C_{12}$ [GPa]  & 61.9${}^g$
    &$61.4 $ & 59.6  & 61.4 \\
    $C_{44}$ [GPa]  & 31.6${}^g$
    &$36.8 $ & 40.4  & 29.7 \\
    $B_\voigt$ [GPa]     & 79.4${}^g$
    &$78.0 $ & 80.1  & 78.1 \\
    $G_\voigt$ [GPa]     & 29.4${}^g$
    &$32.1 $ & 36.5  & 27.0 \\
    $\gamma_{\textrm{sf}}$ [J/m$^2$]    & 0.11--0.21${}^h$
    &$0.142$${}^i$
    & $0.132$ & 0.143  \\
    $\gamma_{\textrm{tsf}}$ [J/m$^2$]   &-  &$0.135$${}^i$
    &  $0.130$ & 0.144  \\
    $T_m$ [K]                 & 935${}^j$
    & 950($\pm 50$)${}^k$
    &918($\pm 5$)   &  898($\pm 5$)  \\
    $\Delta{H}_f$[kJ/mol]     &10.7($\pm0.2$)${}^l$  
    &-   &10.2   &  4.4  \\ 
    $D$   [$10^{-9}$m${}^2$/s]   & 7.2--7.9$^m$   &-   &7.1    &  0.4  \\
    \hline\hline   
  \end{tabular*}
\end{table}

\begin{figure}
  \centering
  \includegraphics[width=0.45\textwidth]{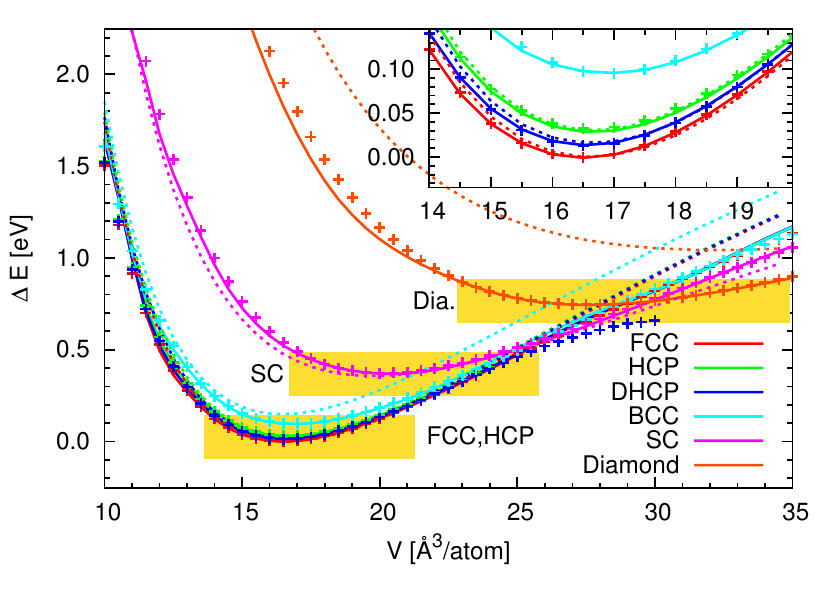}
  \caption{The EOS's of Al. Solid lines, dashed lines, and cross points denote DP, MEAM, and DFT results, respectively. 
    The energies of MEAM are shifted so that the MEAM energy of a stable FCC structure equals to that given by DFT.
  DFT based relaxations fail in some HCP and DHCP structures with a volume per atom larger 30~\AA$^3$, thus the corresponding DFT predictions
  are not shown.
  Yellow bars denote volume ranges in the training data. 
  The BCC and DHCP structures were not explicitly added to the training data.}
  \label{fig:al:01}
\end{figure}

DFT, DP, and MEAM predictions for the equation of state (EOS) of Al are reported in Fig.~\ref{fig:al:01}.
DP reproduces well the DFT results for all the crystalline structures considered here,
i.e., FCC, HCP, double-hexagonal-closed-packed (DHCP), body-centered-cubic (BCC), SC and diamond.
Interestingly, the range of DP accuracy extends well beyond the volume interval that was included in the training data,
which is indicated by the yellow shaded area in the figure.
As shown in the inset of Fig.~\ref{fig:al:01},
the energy difference between FCC and DHCP, and the one between DHCP and HCP is small, only 12~meV/atom and 19~meV/atom, respectively, 
yet DP reproduces accurately the relative stabilities.
The MEAM potential performs well for FCC, HCP, DHCP, and SC, but shows significant deviations from DFT for diamond and BCC.
DP and MEAM predictions for the phonon dispersion relations are compared with experimental results in Fig.~\ref{fig:al:02}.
DP results agree very well with experiment.

The promise of ML potential models is to retain the accuracy of {\it ab initio} molecular dynamics (AIMD) at the cost of FF simulations. 
Therefore, ML potential models can be used to simulate much larger systems for much longer times than possible with AIMD. 
This is illustrated by our calculations for the diffusion coefficient and the radial distribution function (RDF) of the liquid, which were performed on large cells with 4000 atoms with very modest computational resources when using DP.
Thus, the DP model opens opportunities for extending the power of {\it ab initio} methods.

The DP method gives similarly good results for the corresponding properties of pure Mg, which are reported in the SM.

\begin{figure}[t]
  \centering
  \includegraphics[width=0.45\textwidth]{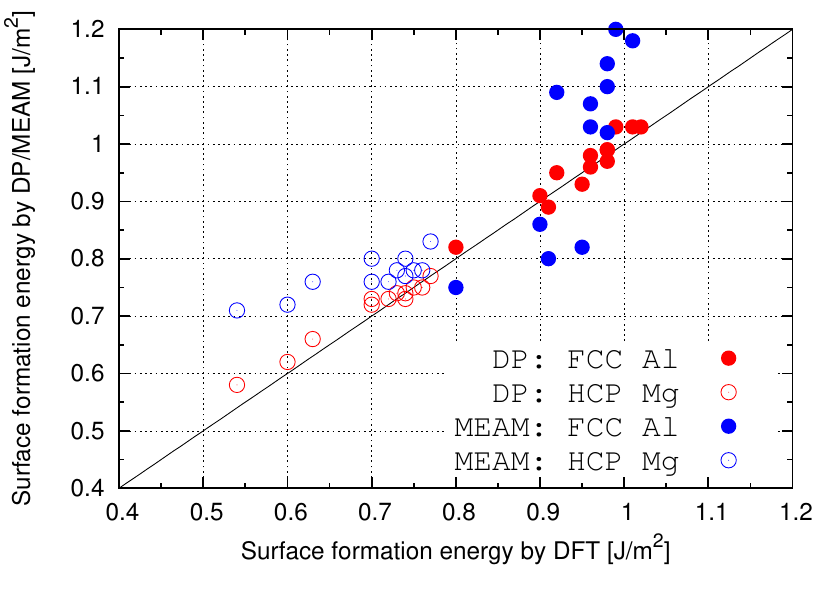}
  \caption{Surface formation energies of Al and Mg.
    All the non-equivalent surfaces with Miller index values smaller than 4 and 3 are investigated for Al and Mg, respectively.
  }
  \label{fig:al.mg:04}
\end{figure}

Finally, we examine the surface formation energy
$E_{\textrm{sf}}((lmn))$,
which describes the energy needed to create a surface with Miller indices $(lmn)$ for a given crystal, and is defined by
$
E_{\textrm{sf}}((lmn)) = \frac{1}{2A}(E_{\textrm{s}}((lmn)) - N_{\textrm{s}} E_0).
$
Here $E_{\textrm{s}}((lmn))$ and $N_{\textrm{s}}$ denote the energy and number of atoms of the relaxed surface structure with Miller indices $(lmn)$.
$A$ denotes the surface area.
We enumerate all the non-equivalent surfaces corresponding to Miller index values smaller than 4 for Al, and smaller than 3 for Mg.
As shown in Fig.~\ref{fig:al.mg:04}, the surfaces formation energies predicted by DP are close to DFT~\cite{tran2016surface}, and those predicted by MEAM are worse in all cases.
We report in detail the values of surface formation energies for Al and Mg in Tables~\ref{tab:al:surf} and \ref{tab:mg:surf}, respectively.

\subsection{Mg-Al Alloys}
\begin{figure*}[t]
\subfloat{\includegraphics[width=0.3\textwidth]{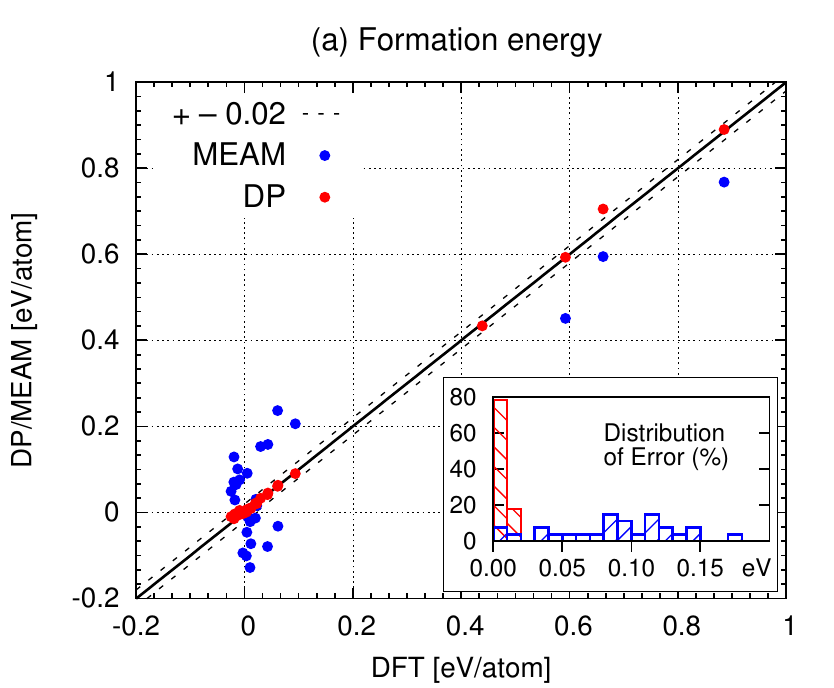}} 
\subfloat{\includegraphics[width=0.3\textwidth]{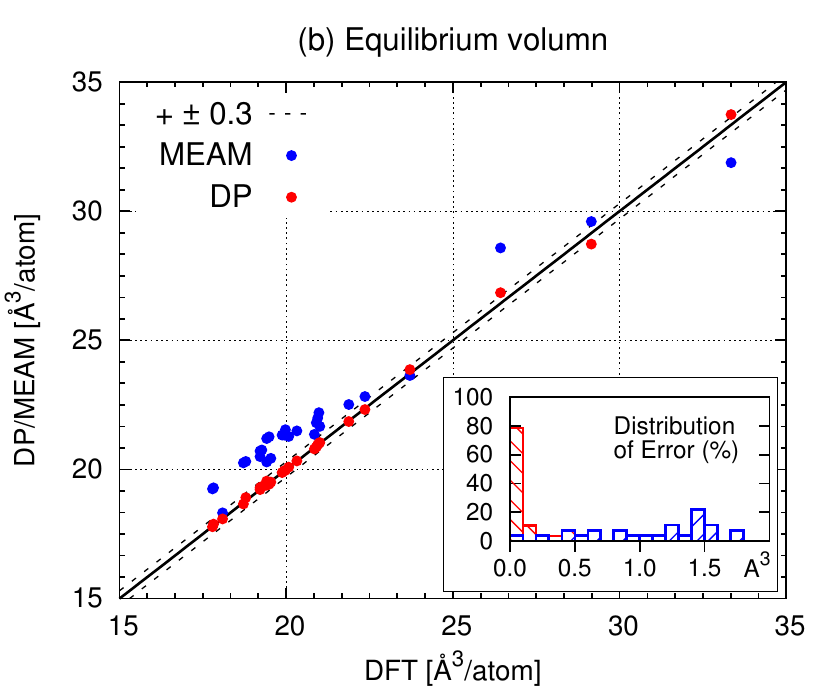}}
\subfloat{\includegraphics[width=0.3\textwidth]{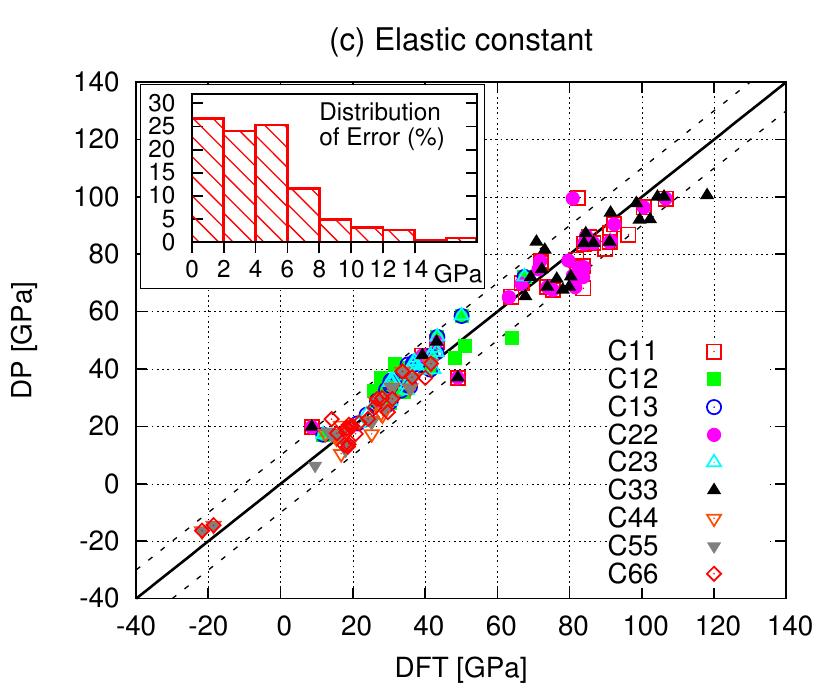}}\\
\subfloat{\includegraphics[width=0.3\textwidth]{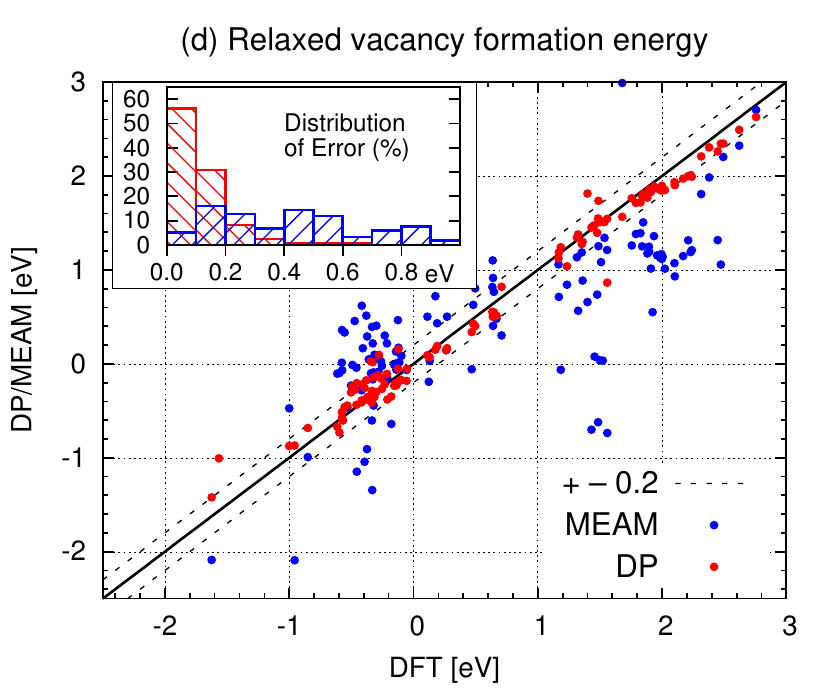}}
\subfloat{\includegraphics[width=0.3\textwidth]{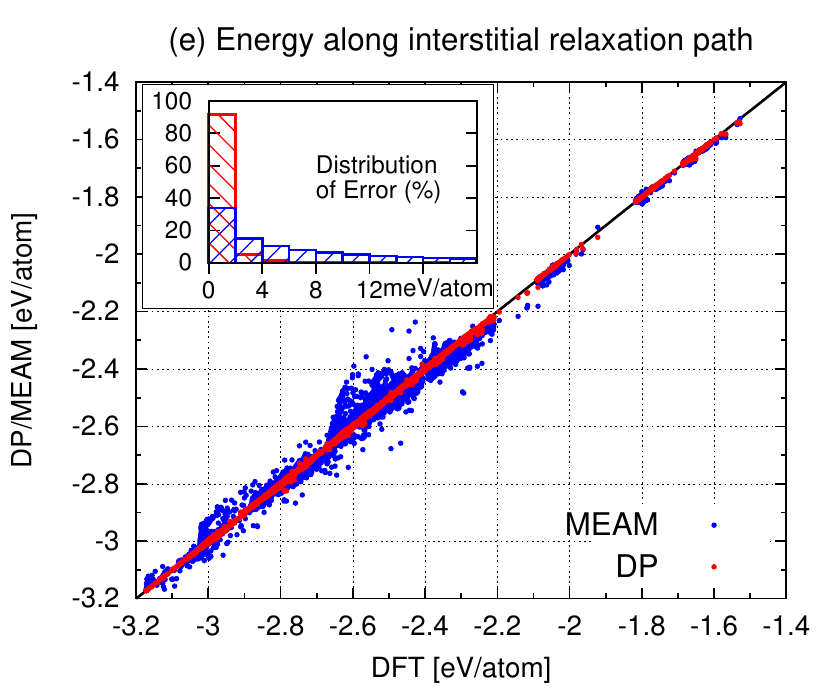}}
\subfloat{\includegraphics[width=0.3\textwidth]{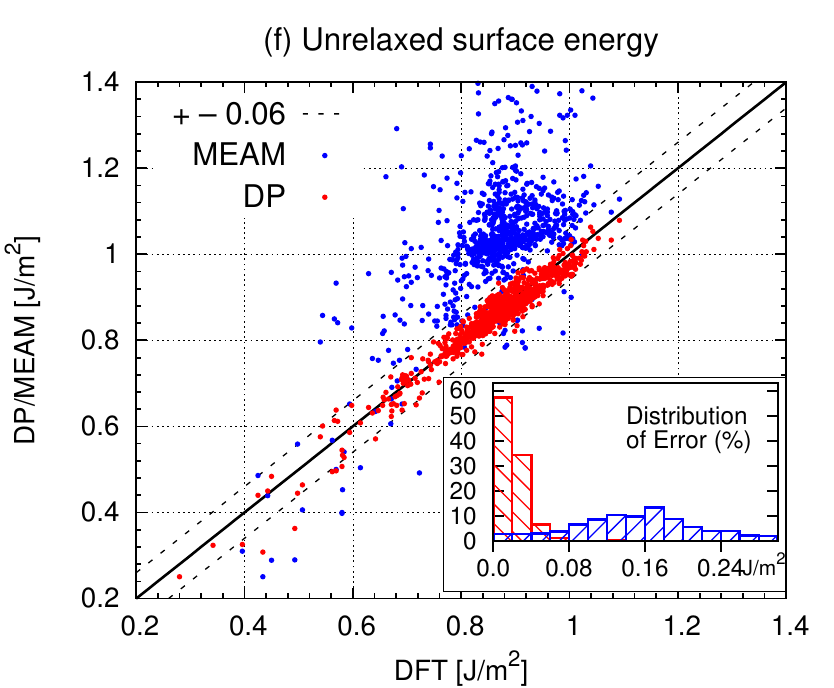}}
\caption{Comparisons of Al-Mg alloy properties predicted by DFT, DP, and MEAM, based on 28 structures in the MP database. 
  (a) 28 formation energies.
  (b) 28 equilibrium volumes per atom.
  (c) 
  225 elastic constants.
  Three structures, mp-1039192, mp-1094664, and mp-12766, are excluded because they resulted unstable under small displacements within DFT.
  (d) 125 relaxed vacancy formation energies.
  Three structures, mp-1039192, mp-1094664, and mp-1038818, are excluded because they resulted unstable under vacancy relaxations within DFT.
  Two DP predictions and three MEAM predictions are outside the range of the plot due to large errors.
  (e) 86,199 energies per atom along the interstitial relaxation pathways within DFT.
  (f) 903 unrelaxed surface energies.
}
\label{fig:mgal}
\end{figure*}

For alloy systems, we adopted the testing scheme introduced in Section I, finding 28 crystalline (ordered) Mg-Al alloy structures in the MP database~\cite{jain2013commentary},
corresponding to relative Mg concentrations ($c_\mg$) ranging from 25\% to 94\%.
Most of these structures were found initially from experiment and were recorded in the inorganic crystal structure database (ICSD)~\cite{hellenbrandt2004inorganic}.
When recorded in the MP database they were further relaxed with DFT.
In Figs.~\ref{fig:mgal} (a-f), we compare predictions of DFT, DP, and MEAM for the 28 alloy structures.
The 6 panels in Fig.~\ref{fig:mgal} report (a) the formation energies, (b) the equilibrium volumes per atom,(c) the elastic constants, (d) the relaxed vacancy formation energies, (e) the total energies per atom along interstitial relaxation pathways, and (f) the unrelaxed surface formation energies.
Notice that only the elastic constants from DP are compared with DFT in Fig.~\ref{fig:mgal} (c).
The corresponding MEAM elastic constants are compared with DFT in Fig.\ref{fig:mgal:02:meam}.

The formation energy of an Mg-Al alloy system is defined as
$$
E_{\textrm{af}} = E_0(c_\mg)
- c_\mg E_{\mg}^0
- (1-c_\mg) E_{\al}^0
$$
where $E_0(c_\mg)$ denotes the equilibrium energy (0~K) per atom of the Mg-Al alloy structure with Mg concentration equal to $c_\mg$,
and $E_{\mg}^0$ and $E_{\al}^0$ denote the equilibrium energies per atom of the corresponding stable crystals of pure Mg and Al at 0~K.
The precise values of the formation energies and equilibrium volumes per atom are reported in Table~\ref{tab:mgal:00}.
To generate the vacancy and interstitial structures, we used supercells that are periodic copies of the MP structures. 
The size of the supercell for each MP structure is reported in Table~\ref{tab:data:dfct-copy}.
We further notice that the interstitial structures are automatically generated based  on 12 MP structures
\footnote{These structures are mp-1038916, mp-1094116, mp-568106, mp-17659, mp-12766, mp-1039141, mp-1094685, mp-2151, mp-1094700, mp-1094970, mp-1016271, mp-1023506}
that are the most stable ones at the corresponding concentrations.
Since most of the interstitial structures are energetically highly unstable,
their relaxation likely ends up with structures that do not represent locally relaxed interstitial point defects, as shown in Fig.~\ref{fig:mgal:04:min}.
In this case, the end structures depend very sensitively on the details of the relaxation.
Therefore, instead of performing independent relaxations within DFT, DP, and MEAM, we compare the predictions of these models for configurations along the DFT relaxation pathways (excluding the initial high energy configurations).

In almost all tested cases, we observe an overall satisfactory agreement between DP predictions and DFT reference results. 
The accuracy of DP is significantly better than that of MEAM.
We stress that the \methodname{} procedure is blind to the alloy structures used to compute the properties reported in Fig.~\ref{fig:mgal}, because these structures were not explicitly included in the training data.
The number of atoms in the unit cell of 6 MP structures is larger than 32, which was the maximum number of atoms in the unit cell of the structures belonging to the training dataset.
This suggests that in the case of Mg-Al alloys the DP model trained with relatively small periodic structures can, to some extent, be used to predict the properties of larger structures.
Some structures tested have little in common with the initial training data. 
Yet the DP model produced satisfactory results, suggesting that it could work for a broader range of materials.

\section{Summary and Outlook}
The DP-GEN scheme is general, practical, and fairly automatic.
To generate the DP model for the Al-Mg system, we did not use any existing DFT database (the MP database was only used for testing),
nor did we use an exhaustive list of possible structures based on physical and chemical considerations.
Instead, we explored the space of configurations using computationally efficient DPMD simulations.
DFT calculations were only performed on a small subset of the configurations that showed large model deviation.
This made possible to progressively improve the DP model.

The DP-GEN scheme is quite flexible.
The three components, training, exploration, and labeling, are highly modularized and can be implemented separately and then recombined.
This makes it easy to incorporate additional functionalities. 
For example, enhanced sampling techniques~\cite{bonati2018silicon} or genetic algorithms~\cite{hajinazar2017stratified} can be incorporated with minimal effort in the exploration module.
We expect that the modular structure of DP-GEN should make possible to use this method to generate models for a variety of important problems, such as finding transition pathways for structural transformations and chemical reactions.
The outcome of DP-GEN include the model and the accumulated data, which could be used for further applications. 
For example, if a rare-earth species is added to the Al-Mg system, one does not need to start the DP-GEN scheme from scratch. 
Instead, one could restart the DP-GEN scheme with the current model and data, and continue with the exploration of the configuration space involving the new species.

Besides alloys dominated by metallic bonding, it would be interesting to use the DP-GEN scheme to study other materials, such as ceramics, polymers, etc., which include different types of bond interactions.
This should be possible because the applicability of DP-GEN relies on three main points: the representability of the model, the validity of the indicator, and the capability of the sampler. 
Several investigations suggest that the first two issues should be relatively independent of the details of the microscopic interactions. 
Indeed, our earlier studies~\cite{zhang2018deep,zhang2018end} indicate that the DP model can represent equally well the PES of systems that differ significantly in their bonding character, such as organic molecules, molecular crystals, hydrogen bonded systems, semiconductors and semimetals. 
In addition, extensive observations by our group show that the DP-GEN indicator, which derives from the variance of the predictions within an ensemble of DNN models, works equally well for different applications~\cite{zhang2018reinforced,zhang2018deepcg}. 
These observations are further supported by recent work by other groups who used closely related indicators in applications to a variety of different systems~\cite{smith2018less,musil2019fast}. 
We are left with the sampler, which may require case specific strategies. 
We are currently investigating this issue in a range of materials, finding that in all cases the search for optimal sampling strategies is facilitated by the modular structure of DP-GEN. 
We will present specific examples in future work.       

Last but not least, one should be aware that DP-GEN scheme may fail in some circumstances. 
We think that this should occur most likely when the sampler and/or the indicator fail. 
For example, the sampler could fail when the  configuration space has high dimensionality and large free energy barriers prevent exploring important configurations. 
In these situations, specifically designed good reaction coordinates might be necessary.
Additional difficulties may be due to the indicator. 
To the best of the authors' knowledge, a rigorous mathematical theory of the indicator is missing. 
A large value of the proposed indicator is only a sufficient, not a necessary, condition for poor performance of a DP model. 
There may be situations in which the physics is poorly described by a model, yet the corresponding ensemble of predictions has small variance. 
We did not face these difficulties in the present investigation but the reader should be aware that systematic validation tests should always be performed before using a DP model to explore new physics.     

\begin{acknowledgements}
The work of L.Z. and W.E is supported in part by Major Program of NNSFC under grant 91130005, ONR grant N00014-13-1-0338 and NSFC grant U1430237.
The work of L.Z. and R.C. is supported in part by the DOE with Award Number DE-SC0019394.
The work of H.W. is supported by the National Science Foundation of China under Grants 11501039, 11871110 and 91530322, and the National Key Research and Development Program of China under Grants 2016YFB0201200 and 2016YFB0201203.
The work of D.Y.L. and H.W. is supported by the Science Challenge Project No.~JCKY2016212A502.
We are grateful for computing time provided in part by 
the  National  Energy  Research  Scientific  Computing  Center  (NERSC),  
the Terascale Infrastructure for Groundbreaking Research in Science and Engineering (TIGRESS) High Performance Computing Center and Visualization Laboratory at Princeton University,
the Special Program for Applied Research on Super Computation of the NSFC-Guangdong Joint Fund under Grant No.~U1501501,
and the Beijing Institute of Big Data Research.
\end{acknowledgements}

\newpage
\beginsupplement

\section{Supplementary Materials}
\subsection{Simulation protocol}
\paragraph{Model}
The smooth edition of the deep potential~\cite{zhang2018end} model is adopted in this work.
The cut-off radius is set to 9~\AA. The $1/r$ terms in the network construction is smoothly switched-off by a cosine shape
function~\cite{zhang2018end} from 2~\AA~to 9~\AA so that the discontinuity due to the cut-off is removed.
The filter (embedding) net is of size $\{25, 50, 100\}$, and the fitting net is of size $\{240, 240, 240\}$. 
A skip connection is built between two neighboring layers, so the architecture of the network is ResNet-like~\cite{he2016deep}.
The Adam stochastic gradient descent method~\cite{Kingma2015adam} is adopted to train the models,
with a learning rate starting at $5.0\times 10^{-4}$ and exponentially decaying to $1.8\times 10^{-8}$ in 400,000 training steps. 
Four models with the same data and training setting,
but different parameter initializations, are trained to estimate the model deviation in the force prediction.
After all the data are collected, the final model is trained with 1,2800,000 training steps.

\paragraph{Exploration} 
Table~\ref{tab:data:02-al}, \ref{tab:data:02-mg}, and Table~\ref{tab:data:03} report the exploration strategy in each iteration for pure Al, pure Mg, and Al-Mg alloy systems, respectively.
During the exploration, if the model deviation of a configuration falls in the range [0.05, 0.15]~eV/\AA\, in the case of pure Al and Al-Mg alloy,
or in range [0.03,0.13]~eV/\AA\, in the case of pure Mg,
then the corresponding configuration is selected for labeling.
The number of atoms in each crystalline structure, the total number of explored and labeled configurations of each crystal structure are reported by Tab.~\ref{tab:data:00}.

\paragraph{Labeling}
The DFT simulation is carried out by the Vienna ab initio simulation package (VASP) version 5.4.4~\cite{kresse1996efficiency,kresse1996efficient},
within the Perdew-Burke-Ernzerhof generalized gradient approximation.
The kinetic energy cutoff for the plane wave expansion is set to 600 eV, and the K-points is set with the Monkhorst-Pack mesh~\cite{monkhorst1976special}
at the spacing $h_k = 0.08~\AA^{-1}$.
The order 1 Methfessel-Paxton smearing method with $\sigma = 0.25$~eV is adopted.
The self-consistent field (SCF) iteration will stop when the total energy and band structure energy differences between
two consecutive steps are smaller than $10^{-6}$~eV.

\begin{table*}
  \centering
  \caption{
    Exploration strategy for the pure Al system.
    For each iteration, we report
    the crystalline structure from which the initial structures are derived,
    the number of DPMD simulations,
    the length of DPMD trajectories,
    the statistical ensemble,
    the temperature of DPMD simulations
    and
    the portion of explored data sent to labeling.
    The pressure is 1~bar  for the NPT ensembles.
  }
  \label{tab:data:02-al}
  \begin{tabular*}{0.99\textwidth}{@{\extracolsep{\fill}} rlrrccc|rlrrccc}
    \hline
    \hline
    Iter.  & Crystal & \#DPMD & Length & Ensemb. & $T$ & Labd. & Iter.  & Crystal & \#DPMD & Length & Ensemb. & $T$ & Labd.\\
           &  &  & [ps] &  & [K] & [\textperthousand] &   &  &  & [ps] &  & [K] & [\textperthousand]\\
    \hline
    0   & FCC, HCP, DIA, SC & 240 & 2 & NVT &  50 &0.33 & 28   & FCC, HCP   & 120 & 2   & NVT &1500 &0.10\\
    1   & FCC, HCP, DIA, SC & 480 & 6 & NVT &  50 &0.03 & 29   & FCC, HCP   & 240 & 6   & NVT &1500 &0.06\\
    2   & FCC, HCP, DIA, SC & 480 & 6 & NVT &  50 &0.03 & 30   & FCC, HCP   & 240 & 6   & NVT &1500 &0.06\\
    3   & FCC, HCP, DIA, SC &1200 & 6 & NVT &  50 &0.01 & 31   & FCC, HCP   & 600 & 6   & NVT &1500 &0.02\\
    4   & FCC, HCP, DIA, SC & 240 & 2 & NVT & 100 &0.08 & 32   & FCC, HCP   & 120 & 2   & NVT &2000 &0.23\\
    5   & FCC, HCP, DIA, SC & 480 & 6 & NVT & 100 &0.02 & 33   & FCC, HCP   & 240 & 6   & NVT &2000 &0.06\\
    6   & FCC, HCP, DIA, SC & 480 & 6 & NVT & 100 &0.03 & 34   & FCC, HCP   & 240 & 6   & NVT &2000 &0.04\\
    7   & FCC, HCP, DIA, SC &1200 & 6 & NVT & 100 &0.01 & 35   & FCC, HCP   & 600 & 6   & NVT &2000 &0.02\\
    8   & FCC, HCP, DIA, SC & 240 & 2 & NVT & 300 &0.19 & 36   & FCC(surf.) & 792 & 0.2 & NVT & 300 &0.69\\
    9   & FCC, HCP, DIA, SC & 480 & 6 & NVT & 300 &0.04 & 37   & FCC(surf.) & 792 & 2   & NVT & 300 &0.08\\
   10   & FCC, HCP, DIA, SC & 480 & 6 & NVT & 300 &0.06 & 38   & FCC(surf.) & 792 & 6   & NVT & 300 &0.03\\
   11   & FCC, HCP, DIA, SC &1200 & 6 & NVT & 300 &0.01 & 39   & FCC(surf.) & 792 & 0.2 & NVT & 900 &0.69\\
   12   & FCC, HCP & 120 & 2 & NVT & 600 &         0.00 & 40   & FCC(surf.) & 792 & 2   & NVT & 900 &0.08\\
   13   & FCC, HCP & 240 & 6 & NVT & 600 &         0.05 & 41   & FCC(surf.) & 792 & 6   & NVT & 900 &0.03\\
   14   & FCC, HCP & 240 & 6 & NVT & 600 &         0.00 & 42   & FCC(surf.) & 792 & 0.2 & NVT &1500 &0.23\\
   15   & FCC, HCP & 600 & 6 & NVT & 600 &         0.00 & 43   & FCC(surf.) & 792 & 2   & NVT &1500 &0.08\\
   16   & FCC, HCP & 120 & 2 & NVT & 900 &         0.00 & 44   & FCC(surf.) & 792 & 6   & NVT &1500 &0.03\\
   17   & FCC, HCP & 240 & 6 & NVT & 900 &         0.06 & 45   & HCP(surf.) & 792 & 0.2 & NVT & 300 &0.00\\
   18   & FCC, HCP & 240 & 6 & NVT & 900 &         0.06 & 46   & HCP(surf.) & 792 &   2 & NVT & 300 &0.00\\
   19   & FCC, HCP & 600 & 6 & NVT & 900 &         0.02 & 47   & HCP(surf.) & 792 &   6 & NVT & 300 &0.00\\
   20   & FCC, HCP & 120 & 2 & NPT & 900 &         0.00 & 48   & HCP(surf.) & 792 & 0.2 & NVT & 900 &0.00\\
   21   & FCC, HCP & 240 & 6 & NPT & 900 &         0.00 & 49   & HCP(surf.) & 792 &   2 & NVT & 900 &0.03\\
   22   & FCC, HCP & 240 & 6 & NPT & 900 &         0.00 & 50   & HCP(surf.) & 792 &   6 & NVT & 900 &0.01\\
   23   & FCC, HCP & 600 & 6 & NPT & 900 &         0.00 & 51   & HCP(surf.) & 792 & 0.2 & NVT & 1500&0.25\\
   24   & FCC, HCP & 120 & 2 & NVT &1200 &         0.12 & 52   & HCP(surf.) & 792 &   2 & NVT & 1500&0.07\\
   25   & FCC, HCP & 240 & 6 & NVT &1200 &         0.06 & 53   & HCP(surf.) & 792 &   6 & NVT & 1500&0.03\\
   26   & FCC, HCP & 240 & 6 & NVT &1200 &         0.06 &      &&&&&&\\
   27   & FCC, HCP & 600 & 6 & NVT &1200 &         0.02 &      &&&&&&\\
    \hline\hline    
  \end{tabular*}
\end{table*}

\begin{table*}
  \centering
  \caption{
    Exploration strategy for the pure Mg system.
    For each iteration, we report
    the crystalline structure from which the initial structures are derived,
    the number of DPMD simulations,
    the length of DPMD trajectories,
    the statistical ensemble,
    the temperature of DPMD simulations
    and
    the portion of explored data sent to labeling.
,The pressure is 1~bar  for the NPT ensembles.
  }
  \label{tab:data:02-mg}
  \begin{tabular*}{0.99\textwidth}{@{\extracolsep{\fill}} rlrrccc|rlrrccc}
    \hline
    \hline
    Iter.  & Crystal & \#DPMD & Length & Ensemb. & $T$ & Labd. & Iter.  & Crystal & \#DPMD & Length & Ensemb. & $T$ & Labd.\\
           &  &  & [ps] &  & [K] & [\textperthousand] &   &  &  & [ps] &  & [K] & [\textperthousand]\\
    \hline
    0   & FCC, HCP, DIA, SC & 240 & 2 & NVT &  50 &0.17 & 28   & FCC, HCP   & 120 & 2   & NVT &1500 &0.00\\
    1   & FCC, HCP, DIA, SC & 480 & 6 & NVT &  50 &0.04 & 29   & FCC, HCP   & 240 & 6   & NVT &1500 &0.04\\
    2   & FCC, HCP, DIA, SC & 480 & 6 & NVT &  50 &0.02 & 30   & FCC, HCP   & 240 & 6   & NVT &1500 &0.03\\
    3   & FCC, HCP, DIA, SC &1200 & 6 & NVT &  50 &0.01 & 31   & FCC, HCP   & 600 & 6   & NVT &1500 &0.02\\
    4   & FCC, HCP, DIA, SC & 240 & 2 & NVT & 100 &0.21 & 32   & FCC, HCP   & 120 & 2   & NVT &2000 &0.00\\
    5   & FCC, HCP, DIA, SC & 480 & 6 & NVT & 100 &0.01 & 33   & FCC, HCP   & 240 & 6   & NVT &2000 &0.06\\
    6   & FCC, HCP, DIA, SC & 480 & 6 & NVT & 100 &0.01 & 34   & FCC, HCP   & 240 & 6   & NVT &2000 &0.06\\
    7   & FCC, HCP, DIA, SC &1200 & 6 & NVT & 100 &0.01 & 35   & FCC, HCP   & 600 & 6   & NVT &2000 &0.02\\
    8   & FCC, HCP, DIA, SC & 240 & 2 & NVT & 300 &0.13 & 36   & FCC(surf.) & 792 & 0.2 & NVT & 300 &0.69\\
    9   & FCC, HCP, DIA, SC & 480 & 6 & NVT & 300 &0.03 & 37   & FCC(surf.) & 792 & 2   & NVT & 300 &0.00\\
   10   & FCC, HCP, DIA, SC & 480 & 6 & NVT & 300 &0.01 & 38   & FCC(surf.) & 792 & 6   & NVT & 300 &0.00\\
   11   & FCC, HCP, DIA, SC &1200 & 6 & NVT & 300 &0.01 & 39   & FCC(surf.) & 792 & 0.2 & NVT & 900 &0.00\\
   12   & FCC, HCP & 120 & 2 & NVT & 600 &         0.33 & 40   & FCC(surf.) & 792 & 2   & NVT & 900 &0.05\\
   13   & FCC, HCP & 240 & 6 & NVT & 600 &         0.06 & 41   & FCC(surf.) & 792 & 6   & NVT & 900 &0.02\\
   14   & FCC, HCP & 240 & 6 & NVT & 600 &         0.00 & 42   & FCC(surf.) & 792 & 0.2 & NVT &1500 &0.10\\
   15   & FCC, HCP & 600 & 6 & NVT & 600 &         0.00 & 43   & FCC(surf.) & 792 & 2   & NVT &1500 &0.07\\
   16   & FCC, HCP & 120 & 2 & NVT & 900 &         0.00 & 44   & FCC(surf.) & 792 & 6   & NVT &1500 &0.03\\
   17   & FCC, HCP & 240 & 6 & NVT & 900 &         0.03 & 45   & HCP(surf.) & 792 & 0.2 & NVT & 300 &0.10\\
   18   & FCC, HCP & 240 & 6 & NVT & 900 &         0.05 & 46   & HCP(surf.) & 792 &   2 & NVT & 300 &0.01\\
   19   & FCC, HCP & 600 & 6 & NVT & 900 &         0.02 & 47   & HCP(surf.) & 792 &   6 & NVT & 300 &0.00\\
   20   & FCC, HCP & 120 & 2 & NPT & 900 &         0.00 & 48   & HCP(surf.) & 792 & 0.2 & NVT & 900 &0.00\\
   21   & FCC, HCP & 240 & 6 & NPT & 900 &         0.00 & 49   & HCP(surf.) & 792 &   2 & NVT & 900 &0.01\\
   22   & FCC, HCP & 240 & 6 & NPT & 900 &         0.00 & 50   & HCP(surf.) & 792 &   6 & NVT & 900 &0.03\\
   23   & FCC, HCP & 600 & 6 & NPT & 900 &         0.00 & 51   & HCP(surf.) & 792 & 0.2 & NVT & 1500&0.55\\
   24   & FCC, HCP & 120 & 2 & NVT &1200 &         0.00 & 52   & HCP(surf.) & 792 &   2 & NVT & 1500&0.04\\
   25   & FCC, HCP & 240 & 6 & NVT &1200 &         0.04 & 53   & HCP(surf.) & 792 &   6 & NVT & 1500&0.03\\
   26   & FCC, HCP & 240 & 6 & NVT &1200 &         0.05 &      &&&&&&\\
   27   & FCC, HCP & 600 & 6 & NVT &1200 &         0.02 &      &&&&&&\\
    \hline\hline    
  \end{tabular*}
\end{table*}

\begin{table*}
  \centering
  \caption{
    Exploration strategy for the Al-Mg alloy system.
    For each iteration, we report
    the crystalline structure from which the initial structures are derived,
    the number of DPMD simulations,
    the length of DPMD trajectories,
    the statistical ensemble,
    the temperature of DPMD simulations
    and
    the portion of explored data sent to labeling.
,The pressure is 1~bar  for the NPT ensembles.
  }
  \label{tab:data:03}
  \begin{tabular*}{0.99\textwidth}{@{\extracolsep{\fill}} rlrrccc|rlrrccc}
    \hline
    \hline
    Iter.  & Crystal & \#DPMD & Length & Ensemb. & $T$ & Labd. & Iter.  & Crystal & \#DPMD & Length & Ensemb. & $T$ & Labd.\\
           &  &  & [ps] &  & [K] & [\textperthousand] &   &  &  & [ps] &  & [K] & [\textperthousand]\\
    \hline
    0   & FCC, HCP, DIA, SC & 408 &0.2& NVT &  50 & 8.15 &   30 &   FCC, HCP   &  552   & 6 & NVT &1500 &0.14\\       
    1   & FCC, HCP, DIA, SC & 816 & 2 & NVT &  50 & 1.68 &   31 &   FCC, HCP   & 1104   & 6 & NVT &1500 &0.20\\      
    2   & FCC, HCP, DIA, SC & 816 & 6 & NVT &  50 & 0.43 &   32 &   FCC, HCP   &  276   & 2 & NVT &2000 &0.00\\      
    3   & FCC, HCP, DIA, SC &1632 & 6 & NVT &  50 & 0.17 &   33 &   FCC, HCP   &  552   & 6 & NVT &2000 &0.33\\      
    4   & FCC, HCP, DIA, SC & 408 &0.2& NVT & 100 & 3.24 &   34 &   FCC, HCP   &  552   & 6 & NVT &2000 &0.31\\      
    5   & FCC, HCP, DIA, SC & 816 & 2 & NVT & 100 & 0.41 &   35 &   FCC, HCP   & 1104   & 6 & NVT &2000 &0.17\\      
    6   & FCC, HCP, DIA, SC & 816 & 6 & NVT & 100 & 0.12 &   36 &   FCC(surf.) & 11,616 &0.2& NVT &300  &0.69\\
    7   & FCC, HCP, DIA, SC &1632 & 6 & NVT & 100 & 0.07 &   37 &   FCC(surf.) & 11,616 & 2 & NVT &300  &0.07\\
    8   & FCC, HCP, DIA, SC & 408 &0.2& NVT & 300 & 1.36 &   38 &   FCC(surf.) & 11,616 & 6 & NVT &300  &0.00\\
    9   & FCC, HCP, DIA, SC & 816 & 2 & NVT & 300 & 0.82 &   39 &   FCC(surf.) & 11,616 & 6 & NVT &300  &0.01\\
   10   & FCC, HCP, DIA, SC & 816 & 6 & NVT & 300 & 0.17 &   40 &   FCC(surf.) & 11,616 &0.2& NVT &900  &0.13\\
   11   & FCC, HCP, DIA, SC &1632 & 6 & NVT & 300 & 0.05 &   41 &   FCC(surf.) & 11,616 & 2 & NVT &900  &0.01\\
   12   & FCC, HCP & 276 & 2 & NVT & 600 &          0.29 &   42 &   FCC(surf.) & 11,616 & 6 & NVT &900  &0.02\\
   13   & FCC, HCP & 552 & 6 & NVT & 600 &          0.26 &   43 &   FCC(surf.) & 11,616 & 6 & NVT &900  &0.00\\
   14   & FCC, HCP & 552 & 6 & NVT & 600 &          0.35 &   44 &   FCC(surf.) & 11,616 &0.2& NVT &1500 &0.01\\
   15   & FCC, HCP &1104 & 6 & NVT & 600 &          0.03 &   45 &   FCC(surf.) & 11,616 & 2 & NVT &1500 &0.01\\
   16   & FCC, HCP & 276 & 2 & NVT & 900 &          0.52 &   46 &   FCC(surf.) & 11,616 & 6 & NVT &1500 &0.02\\
   17   & FCC, HCP & 552 & 6 & NVT & 900 &          0.50 &   47 &   FCC(surf.) & 11,616 & 6 & NVT &1500 &0.03\\
   18   & FCC, HCP & 552 & 6 & NVT & 900 &          0.08 &   48 &   HCP(surf.) & 11,616 &0.2& NVT &300  &0.00\\
   19   & FCC, HCP &1104 & 6 & NVT & 900 &          0.00 &   49 &   HCP(surf.) & 11,616 & 2 & NVT &300  &0.00\\
   20   & FCC, HCP & 276 & 2 & NPT & 900 &          0.00 &   50 &   HCP(surf.) & 11,616 & 6 & NVT &300  &0.00\\
   21   & FCC, HCP & 552 & 6 & NPT & 900 &          0.00 &   51 &   HCP(surf.) & 11,616 & 6 & NVT &300  &0.00\\
   22   & FCC, HCP & 552 & 6 & NPT & 900 &          0.01 &   52 &   HCP(surf.) & 11,616 &0.2& NVT &900  &0.00\\
   23   & FCC, HCP &1104 & 6 & NPT & 900 &          0.00 &   53 &   HCP(surf.) & 11,616 & 2 & NVT &900  &0.00\\
   24   & FCC, HCP & 276 & 2 & NVT &1200 &          0.04 &   54 &   HCP(surf.) & 11,616 & 6 & NVT &900  &0.00\\
   25   & FCC, HCP & 552 & 6 & NVT &1200 &          0.47 &   55 &   HCP(surf.) & 11,616 & 6 & NVT &900  &0.00\\
   26   & FCC, HCP & 552 & 6 & NVT &1200 &          0.03 &   56 &   HCP(surf.) & 11,616 &0.2& NVT &1500 &0.00\\
   27   & FCC, HCP &1104 & 6 & NVT &1200 &          0.21 &   57 &   HCP(surf.) & 11,616 & 2 & NVT &1500 &0.00\\
   28   & FCC, HCP & 276 & 2 & NVT &1500 &          0.00 &   58 &   HCP(surf.) & 11,616 & 6 & NVT &1500 &0.01\\
   29   & FCC, HCP & 552 & 6 & NVT &1500 &          0.35 &   59 &   HCP(surf.) & 11,616 & 6 & NVT &1500 &0.02\\
    \hline\hline    
  \end{tabular*}
\end{table*}

\begin{table*}
  \centering
  \caption{Number of explored and labeled data by the \methodname{} scheme for pure Al, pure Mg, and Mg-Al alloy systems.  }
  \label{tab:data:00}
  \begin{tabular*}{0.99\textwidth}{@{\extracolsep{\fill}} clrrrrrrr}
    \hline\hline    
    \multicolumn{3}{c}{{Systems}} &
    \multicolumn{2}{c}{Al} &
    \multicolumn{2}{c}{Mg} &
    \multicolumn{2}{c}{Al-Mg alloys} \\\cline{1-3}\cline{4-5}\cline{6-7}\cline{8-9}
    Type& Crystal & \#atom
    & \#Explored & \#Labeled
    & \#Explored & \#Labeled
    & \#Explored & \#Labeled
    \\
    \hline
    \multirow{4}{*}{Bulk}
    &FCC       &32 &  15,174,000  & 1,326    & 15,174,000  & 860   &39,266,460   &7,313  \\
    &HCP       &16 &  15,174,000  & 908      & 15,174,000  & 760   &18,999,900   &2,461  \\
    &Diamond   &16 &  5,058,000   & 1,026    & 5,058,000   & 543   &5,451,300    &2,607  \\
    &SC        & 8 &  5,058,000   & 713      & 5,058,000   & 234   &2,543,940    &667    \\\hline
    \multirow{6}{*}{Surface}
    &FCC (100) &12 &  3,270,960   & 728      & 3,270,960   & 251   &62,203,680   &1,131  \\
    &FCC (110) &16\footnote{Pure Al\label{first_footnote}},20\footnote{Mg and Al-Mg alloy\label{2nd_footnote}} &  3,270,960   & 838      & 3,270,960   & 353   &10,744,2720  &2,435  \\
    &FCC (111) &12 &  3,270,960   & 544      & 3,270,960   & 230   &62,203,680   &1,160  \\
    &HCP ($0001$)     &12 & 3,270,960 & 39   & 3,270,960   & 109   &62,203,680   &176    \\
    &HCP ($10\bar10$) &12 & 3,270,960 & 74   & 3,270,960   & 167   &62,203,680   &203    \\
    &HCP ($11\bar20$) &16\footref{first_footnote},20\footref{2nd_footnote} & 3,270,960 & 293  & 3,270,960   & 182   &107,442,720  &501    \\\hline
    sum & & & 60,089,760 & 6,489          & 60,089,760 & 3,689 &529,961,760  &18,654 \\
    \hline\hline    
  \end{tabular*}
\end{table*}

\begin{table*}
  \centering
  \caption{The size of supercells used for generating the vacancy and interstitial structures. 
  }
  \label{tab:data:dfct-copy}
  \begin{tabular*}{0.95\textwidth}{@{\extracolsep{\fill}} lcc|lcc|lcc}
    \hline\hline    
    Mat.Proj.ID& \# copies & $N$ & Mat.Proj.ID& \# copies & $N$ & Mat.Proj.ID& \# copies & $N$ \\
    \hline
    mp-1016233& $2\times3\times2$&48 &    mp-1039141& $3\times3\times3$&54 &     mp-1094700& $1\times1\times1$&58 \\
    mp-1016271& $2\times2\times2$&64 &    mp-1039180& $1\times1\times1$&16 &     mp-1094961& $2\times2\times1$&32 \\
    mp-1023506& $2\times2\times1$&64 &    mp-1039192& $2\times2\times2$&48 &     mp-1094970& $2\times2\times2$&64 \\
    mp-1038779& $3\times3\times1$&54 &    mp-1094116& $1\times1\times1$&12 &     mp-1094987& $2\times2\times2$&32 \\
    mp-1038818& $2\times2\times2$&64 &    mp-1094664& $2\times2\times2$&32 &     mp-12766  & $1\times1\times1$&27 \\
    mp-1038916& $2\times2\times1$&32 &    mp-1094666& $1\times1\times1$&16 &     mp-17659  & $1\times1\times1$&53 \\
    mp-1038934& $3\times3\times3$&54 &    mp-1094683& $1\times1\times1$&58 &     mp-2151   & $1\times1\times1$&29 \\
    mp-1039010& $2\times2\times2$&64 &    mp-1094685& $1\times1\times1$&58 &     mp-568106 & $1\times1\times1$&108\\
    mp-1039019& $3\times3\times3$&54 &    mp-1094692& $1\times1\times1$&87 &     mp-978271 & $2\times2\times2$&32 \\
    mp-1039119& $2\times2\times2$&64 &    &&&&&\\
    \hline\hline    
  \end{tabular*}
\end{table*}

\subsection{Additional simulation results}

\begin{table*}[]
  \centering
  \caption{Surface formation energies of FCC Al.}
  \label{tab:al:surf}
  \begin{tabular*}{0.48\textwidth}{@{\extracolsep{\fill}} cccc}
    \hline
    \hline
    Miller & DFT\cite{tran2016surface} & DP & MEAM \\
    indices& [J/m$^2$]& [J/m$^2$]& [J/m$^2$] \\
    \hline
    (100) & 0.92 &   0.95 &  1.09 \\
    (110) & 0.98 &   0.99 &  1.14 \\
    (111) & 0.80 &   0.82 &  0.75 \\
    (210) & 1.02 &   1.03 &  1.21 \\
    (211) & 0.98 &   0.99 &  1.02 \\
    (221) & 0.95 &   0.93 &  0.82 \\
    (310) & 0.99 &   1.03 &  1.20 \\
    (311) & 0.98 &   0.97 &  1.10 \\
    (320) & 1.01 &   1.03 &  1.18 \\
    (321) & 0.96 &   0.98 &  1.07 \\
    (322) & 0.90 &   0.91 &  0.86 \\
    (331) & 0.96 &   0.96 &  1.03 \\
    (332) & 0.91 &   0.89 &  0.80 \\
    \hline
    \hline
  \end{tabular*}
\end{table*}

\begin{table*}[]
  \centering
  \caption{Surface formation energies of HCP Mg.}
  \label{tab:mg:surf}
  \begin{tabular*}{0.48\textwidth}{@{\extracolsep{\fill}} cccc}
    \hline
    \hline
    Miller & DFT\cite{tran2016surface} & DP & MEAM \\
    indices& [J/m$^2$]& [J/m$^2$]& [J/m$^2$] \\
    \hline
($0001  $)        &0.54 &0.58   &0.71\\
($10\bar10  $)&0.60 &0.62   &0.72\\
($10\bar11  $)&0.63 &0.66   &0.76\\
($10\bar12  $)&0.70 &0.73   &0.80\\
($11\bar20  $)&0.72 &0.73   &0.76\\
($11\bar21  $)&0.76 &0.75   &0.78\\
($20\bar21  $)&0.77 &0.77   &0.83\\
($2\bar1\bar12  $)&0.74 &0.73   &0.77\\
($21\bar30  $)&0.70 &0.72   &0.76\\
($21\bar31  $)&0.73 &0.74   &0.78\\
($21\bar32  $)&0.74 &0.74   &0.80\\
($22\bar41  $)&0.75 &0.75   &0.78\\
    \hline                   
    \hline
  \end{tabular*}
\end{table*}

\begin{table*}[]
  \centering
  \caption{Equilibrium properties of Mg.}
  \label{tab:mg:00}
  \begin{tabular*}{0.48\textwidth}{@{\extracolsep{\fill}} lrrrr}
    \hline
    \hline
Mg & Exp. & DFT  & DP & MEAM  \\\hline
$E_0$ [eV/atom] & $-1.51$\footnote{Reference~\cite{kittel2004introduction}\label{mge0}}   &$-1.489$ & $-1.487$ &  $-1.510$ \\
$V_0$ [\AA$^3$/atom] & $23.24$\footnote{Reference~\cite{von1957lattice}, 300K}   &22.89  &   22.83 &  23.04  \\
$E_{\textrm{vf}}$ [eV] & 0.79\footnote{Reference~\cite{tzanetakis1976formation}}     & 0.79 & 0.67 & 0.87 \\
    \multirow{4}{*}{$E_{\textrm{if}}$[eV]} &  \multirow{4}{*}{-} & 2.16 & 2.17 & 1.76 \\
    &&2.18 & 2.14 & 1.76 \\
    &&2.19 & 2.23 & 1.56 \\
    &&2.16 & 2.14 & 1.56\\
$C_{11}$ [GPa] & 63.5\footnote{Reference~\cite{slutsky1957elastic}\label{mgela}} & 61.0& 59.9& 59.7 \\
$C_{33}$ [GPa] & 66.5\footref{mgela} & 64.5& 71.3& 60.7 \\
$C_{12}$ [GPa] & 25.9\footref{mgela} & 28.2& 26.1& 23.2 \\
$C_{13}$ [GPa] & 21.7\footref{mgela} & 20.7& 19.2& 23.6 \\
$C_{44}$ [GPa] & 18.4\footref{mgela} & 16.0& 16.7& 16.7 \\
$C_{66}$ [GPa] & 18.8\footref{mgela} & 18.4& 16.8& 18.2 \\
$B_\voigt$ [GPa]  & 36.9\footref{mgela} & 36.3 & 35.6 & 35.7 \\
$G_\voigt$ [GPa]  & 19.4\footref{mgela} & 17.9 & 18.6 & 17.6 \\
    $\gamma_{\textrm{sf}}$ [J/m$^2$]  & 0.078\footnote{Reference~\cite{sastry1969stacking}}  & 0.0339\footnote{Reference~\cite{zhang2017first}\label{refmg}}
         & 0.0384 &  0.0218 \\
    $\gamma_{\textrm{tsf}}$ [J/m$^2$] & -  & 0.0414\footref{refmg}  & 0.0388 &  0.0219 \\
     $T_m$ [K]                 & 922 &  - & 870  & 765   \\
     $\Delta{H}_f$[kJ/mol]     &9.0($\pm0.2$)${}^l$  &-   &8.7   &  9.1  \\ 
    \hline
    \hline
  \end{tabular*}
\end{table*}

\begin{figure}
  \centering
  \includegraphics[width=0.45\textwidth]{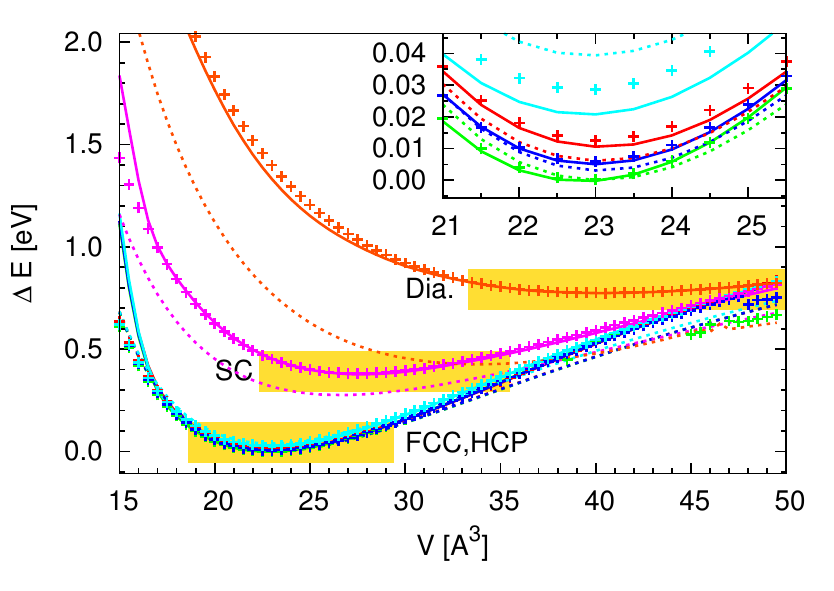}
  \caption{Equation of states of Mg.
    Solid lines denotes DP results. Dashed lines denote MEAM results. Cross points denote DFT results. 
    The energies of DP and MEAM are shifted so that the MEAM energy of a stable FCC structure equils to that given by DFT.
    The DFT relaxations fail in some HCP and DHCP structures with atomic volume larger 44~\AA$^3$, thus the EOSs of HCP and DHCP beyond 44~\AA$^3$
    are not shown.
  }
  \label{fig:mg:01}
\end{figure}

\begin{figure}
  \centering
  \includegraphics[width=0.45\textwidth]{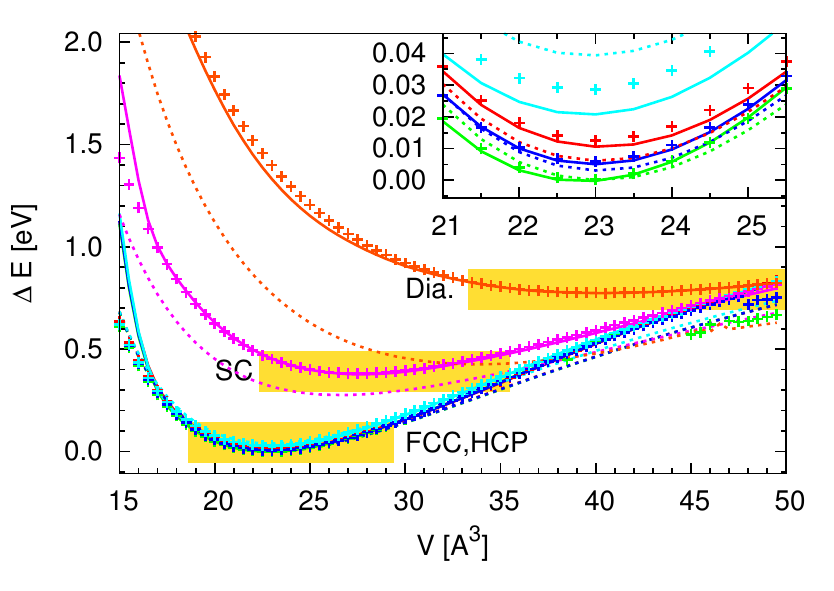}
  \caption{The RDFs of liquid Mg at $P=1$~bar and temperatures $T = 953$~K.
    The DP and MEAM predictions are compared with the experimental data taken from~\cite{waseda1980structure}.
    The inserted plot is the zoom-in of RDFs in range $3.5$\AA$\leq r \leq 7$\AA.
  }
  \label{fig:mg:02}
\end{figure}

\begin{figure}
  \centering
  \includegraphics[width=0.48\textwidth]{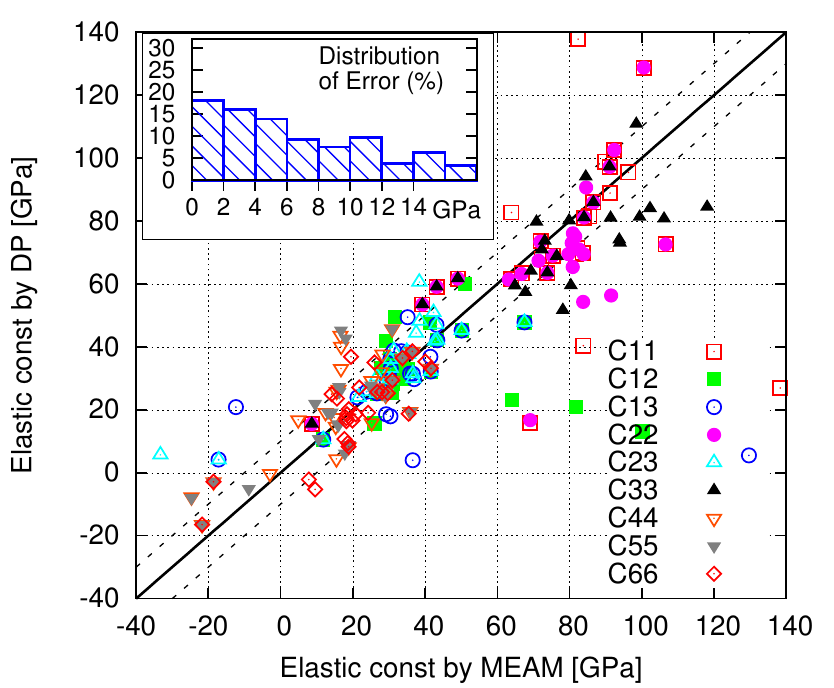}
  \caption{Elastic constants computed by MEAM compared with those computed by DFT.}
  \label{fig:mgal:02:meam}
\end{figure}

\begin{figure}
  \centering
  \includegraphics[width=0.48\textwidth]{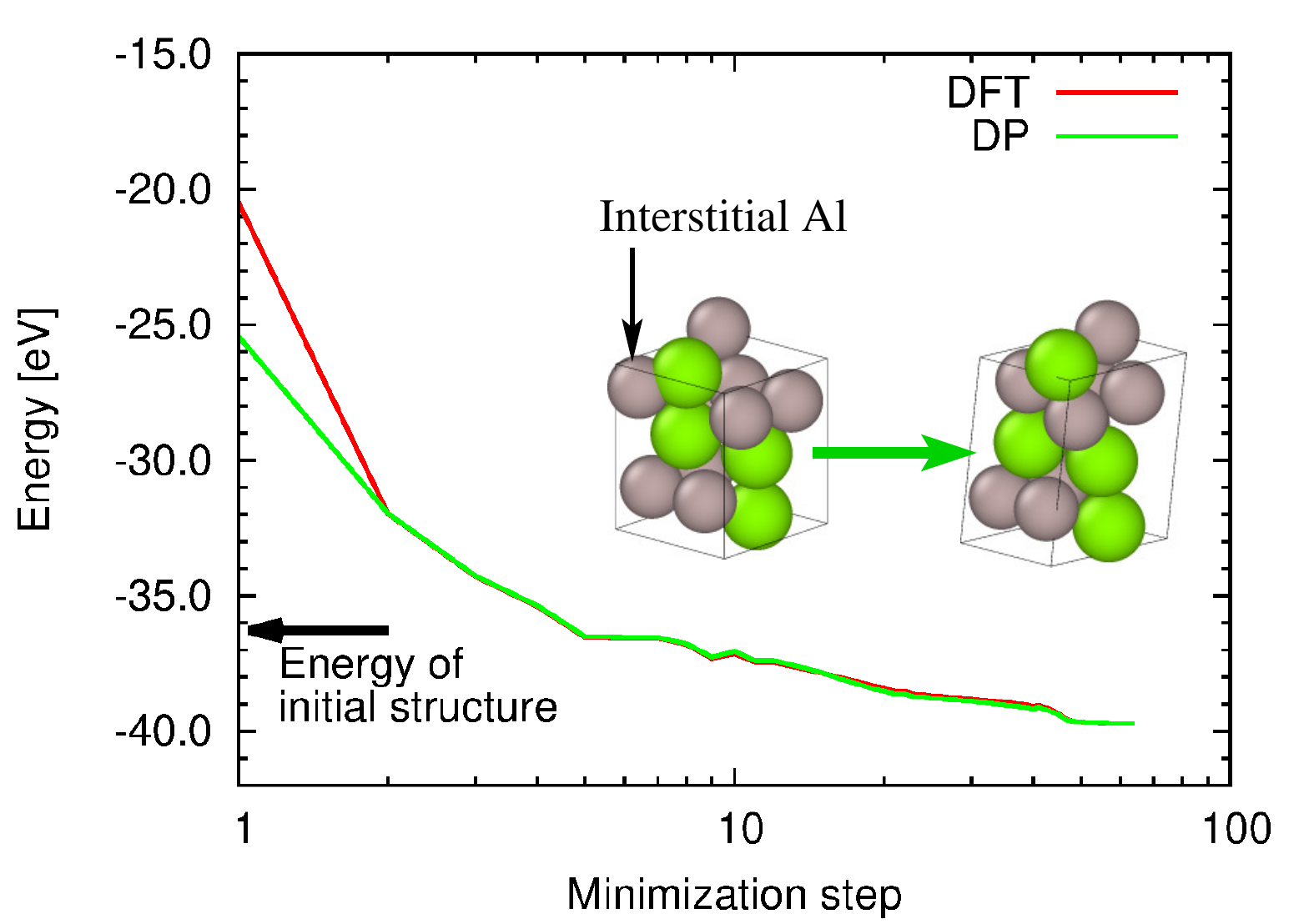}
  \caption{Comparison of energies predicted by DFT and DP along the Al interstitial relaxation pathway within DFT.
    The interstitial structure is generated from mp-1094116.
    The insert shows the initial and relaxed interstitial configurations.
  }
  \label{fig:mgal:04:min}
\end{figure}

\begin{table*}
  \centering
  \caption{The formation energy and volume per atom of Mg-Al alloys, displayed in ascending order of Mg concentration and descending order of structure stability.
    The numbers in the parenthese give the difference between the DP/MEAM method and the DFT calculation.
  }
  \label{tab:mgal:00}
  \begin{tabular*}{\textwidth}{@{\extracolsep{\fill}} lcrrr@{\hskip -.1cm}rr@{\hskip -.1cm}r@{\hskip .2cm}rr@{\hskip -.1cm}rr@{\hskip -.1cm}r}
    \hline\hline
    Mat.Proj.ID & $N$ & $c_{\textrm{Mg}}$ & $E^\dft_\aform$  & \multicolumn{2}{c}{$E^\deep_\aform$}  & \multicolumn{2}{c}{$E^\meam_\aform$} & $V^\dft_0$ & \multicolumn{2}{c}{$V^\deep_0$} & \multicolumn{2}{c}{$ V^\meam_0$} \\
                &     &      &           [eV/atom] & \multicolumn{2}{c}{[eV/atom]}            & \multicolumn{2}{c}{[eV/atom]}           & [\AA$^3$] & \multicolumn{2}{c}{[\AA$^3$]}  & \multicolumn{2}{c}{[\AA$^3$]}   \\
    \hline    
mp-1038916&   8&  25.0\%&  $  0.005$&  $  0.007$& ( 2)&  $  0.091$& ( 86)&  17.79&  17.78& ($ 0.1$\%)&  19.25& ($ 8.2$\%) \\ 
mp-1039119&   8&  25.0\%&  $  0.020$&  $  0.021$& ( 1)&  $ -0.013$& ( 33)&  17.82&  17.89& ($ 0.4$\%)&  19.29& ($ 8.3$\%) \\ 
mp-1039180&  16&  25.0\%&  $  0.094$&  $  0.090$& ( 4)&  $  0.206$& (112)&  18.09&  18.09& ($ 0.0$\%)&  18.32& ($ 1.3$\%) \\ 
mp-1094116&  12&  33.3\%&  $ -0.013$&  $ -0.007$& ( 6)&  $  0.101$& (114)&  18.79&  18.91& ($ 0.6$\%)&  20.31& ($ 8.1$\%) \\ 
mp-1039192&   6&  33.3\%&  $  0.662$&  $  0.705$& (43)&  $  0.595$& ( 67)&  26.43&  26.84& ($ 1.6$\%)&  28.58& ($ 8.1$\%) \\ 
mp-568106&  108&  38.0\%&  $  0.023$&  $  0.021$& ( 2)&  $  0.016$& (  7)&  18.72&  18.66& ($ 0.3$\%)&  20.26& ($ 8.2$\%) \\ 
mp-17659&    53&  43.4\%&  $ -0.019$&  $ -0.014$& ( 5)&  $  0.129$& (149)&  19.27&  19.34& ($ 0.4$\%)&  20.76& ($ 7.7$\%) \\ 
mp-12766&    27&  48.1\%&  $  0.061$&  $  0.061$& ( 0)&  $  0.237$& (175)&  19.54&  19.51& ($ 0.2$\%)&  20.43& ($ 4.6$\%) \\ 
mp-1039141&   2&  50.0\%&  $ -0.003$&  $ -0.003$& ( 0)&  $ -0.094$& ( 91)&  19.22&  19.22& ($ 0.0$\%)&  20.71& ($ 7.8$\%) \\ 
mp-1094987&   4&  50.0\%&  $  0.010$&  $  0.010$& ( 0)&  $ -0.128$& (138)&  19.22&  19.30& ($ 0.4$\%)&  20.50& ($ 6.7$\%) \\ 
mp-1038779&   6&  50.0\%&  $  0.030$&  $  0.033$& ( 3)&  $  0.153$& (124)&  19.49&  19.43& ($ 0.3$\%)&  21.27& ($ 9.1$\%) \\ 
mp-1039019&   2&  50.0\%&  $  0.043$&  $  0.045$& ( 2)&  $  0.158$& (115)&  19.41&  19.39& ($ 0.1$\%)&  21.20& ($ 9.2$\%) \\ 
mp-1038934&   2&  50.0\%&  $  0.062$&  $  0.063$& ( 1)&  $ -0.032$& ( 93)&  19.41&  19.55& ($ 0.7$\%)&  20.29& ($ 4.5$\%) \\ 
mp-1039010&   8&  50.0\%&  $  0.439$&  $  0.433$& ( 5)&  $  0.230$& (209)&  23.71&  23.87& ($ 0.7$\%)&  23.64& ($ 0.3$\%) \\ 
mp-1038818&   8&  50.0\%&  $  0.885$&  $  0.890$& ( 5)&  $  0.768$& (118)&  33.34&  33.74& ($ 1.2$\%)&  31.88& ($ 4.4$\%) \\ 
mp-1094664&   4&  50.0\%&  $  0.592$&  $  0.593$& ( 0)&  $  0.451$& (141)&  29.15&  28.73& ($ 1.4$\%)&  29.60& ($ 1.6$\%) \\ 
mp-1094685&  58&  55.2\%&  $ -0.019$&  $ -0.007$& (13)&  $  0.071$& ( 90)&  19.88&  19.88& ($ 0.0$\%)&  21.33& ($ 7.3$\%) \\ 
mp-1094692&  87&  55.2\%&  $ -0.015$&  $ -0.003$& (12)&  $  0.065$& ( 81)&  19.94&  19.95& ($ 0.0$\%)&  21.33& ($ 7.0$\%) \\ 
mp-1094683&  58&  55.2\%&  $ -0.009$&  $  0.005$& (14)&  $  0.076$& ( 85)&  19.98&  19.97& ($ 0.0$\%)&  21.54& ($ 7.8$\%) \\ 
mp-2151&     29&  58.6\%&  $ -0.025$&  $ -0.010$& (15)&  $  0.049$& ( 74)&  20.07&  20.09& ($ 0.1$\%)&  21.28& ($ 6.0$\%) \\ 
mp-1094700&  58&  62.1\%&  $ -0.018$&  $ -0.003$& (15)&  $  0.029$& ( 47)&  20.32&  20.33& ($ 0.1$\%)&  21.50& ($ 5.8$\%) \\ 
mp-1094970&   8&  75.0\%&  $  0.004$&  $  0.001$& ( 3)&  $ -0.101$& (104)&  20.90&  20.87& ($ 0.2$\%)&  21.81& ($ 4.3$\%) \\ 
mp-978271&    4&  75.0\%&  $  0.005$&  $  0.003$& ( 2)&  $ -0.046$& ( 50)&  20.95&  20.96& ($ 0.1$\%)&  21.99& ($ 5.0$\%) \\ 
mp-1094961&   8&  75.0\%&  $  0.012$&  $  0.009$& ( 2)&  $ -0.072$& ( 84)&  21.01&  21.04& ($ 0.2$\%)&  21.67& ($ 3.1$\%) \\ 
mp-1016233&   4&  75.0\%&  $  0.021$&  $  0.021$& ( 0)&  $  0.031$& (  9)&  20.98&  21.01& ($ 0.1$\%)&  22.20& ($ 5.8$\%) \\ 
mp-1094666&  16&  75.0\%&  $  0.043$&  $  0.042$& ( 1)&  $ -0.079$& (122)&  20.85&  20.80& ($ 0.2$\%)&  21.36& ($ 2.5$\%) \\ 
mp-1016271&   8&  87.5\%&  $  0.011$&  $  0.010$& ( 1)&  $ -0.021$& ( 32)&  21.88&  21.86& ($ 0.1$\%)&  22.51& ($ 2.9$\%) \\ 
mp-1023506&  16&  93.8\%&  $  0.006$&  $  0.005$& ( 0)&  $ -0.010$& ( 15)&  22.36&  22.32& ($ 0.2$\%)&  22.82& ($ 2.1$\%) \\ 
    \hline\hline    
  \end{tabular*}
\end{table*}


\end{document}